\documentclass[conference]{IEEEtran}
\IEEEoverridecommandlockouts
\usepackage{cite}
\usepackage{amsmath,amssymb,amsfonts}
\usepackage{algorithmic}
\usepackage{graphicx}
\usepackage{textcomp}
\usepackage{xcolor}
\usepackage[hyphens]{url}
\usepackage{fancyhdr}
\usepackage{hyperref}

\makeatletter
\let\MYcaption\@makecaption
\makeatother

\usepackage{subcaption}

\makeatletter
\let\@makecaption\MYcaption
\makeatother

\usepackage{enumitem}
\usepackage{pifont}
\usepackage{xcolor}
\usepackage{multirow}
\usepackage{balance}
\definecolor{darkgreen}{rgb}{0.0, 0.5, 0.0}
\definecolor{darkyellow}{RGB}{194, 162, 4}

\newcommand{\redtext}{\textcolor{red}}
\newcommand{\greentext}{\textcolor{darkgreen}}
\newcommand{\orangetext}{\textcolor{orange}}
\newcommand{\yellowtext}{\textcolor{darkyellow}}

\usepackage{tikz}
\newcommand*\circled[1]{\tikz[baseline=(char.base)]{
            \node[shape=circle,draw,inner sep=0.5pt] (char) {#1};}}

\title{Pinball: A Cryogenic Predecoder for Surface Code Decoding Under Circuit-Level Noise}

\newcommand\paperauthors{Alexander Knapen$^{\dagger*}$, Guanchen Tao$^{\dagger*}$, Jacob Mack$^\dagger$, Tomas Bruno$^\dagger$, Mehdi Saligane$^\ddagger$, Dennis Sylvester$^\dagger$,\\ Qirui Zhang$^\dagger$, and Gokul Subramanian Ravi$^\dagger$}
\newcommand\affiliations{$^\dagger$University of Michigan, Ann Arbor; $^\ddagger$Brown University}
\newcommand\emails{Email(s): \{aknapen, guanchen, jmackmi, tbruno\}@umich.edu, mehdi\_saligane@brown.edu,\\\{dmcs, qiruizh, gsravi\}@umich.edu}

\author{
    \IEEEauthorblockN{\paperauthors{}}
      \IEEEauthorblockA{
        \affiliations{}\\
        \emails{}
      }
}

\begin{document}

\maketitle

\makeatletter
\def\blfootnote{\gdef\@thefnmark{}\@footnotetext}
\makeatother

\blfootnote{*These authors contributed equally to this work.}
\begin{abstract}
Scaling fault tolerant quantum computers, especially cryogenic systems based on the surface code, to millions of qubits is very challenging due to poorly-scaling data processing and power consumption overheads. One key challenge is the design of decoders for real-time quantum error correction (QEC), which demands high data rates for error processing; this is particularly apparent in systems with cryogenic qubits and room temperature (RT) decoders. In response, cryogenic predecoding using lightweight logic has been proposed to handle common, sparse errors within the cryogenic domain. However, prior work only accounts for a subset of the error sources present in real-world quantum systems with limited accuracy, often degrading performance below a useful level in practical scenarios. Furthermore, prior reliance on SFQ logic precludes detailed architecture-technology co-optimization.

To address these shortcomings, this paper introduces Pinball\footnote{Source code available at: https://github.com/aknapen/Pinball}, a comprehensive design in cryogenic CMOS of a QEC predecoder for the surface code, tailored to realistic, circuit-level noise. By accounting for error generation and propagation through QEC circuits, our design achieves higher predecoding accuracy, outperforming logical error rates of the current state-of-the-art cryogenic predecoder by nearly \emph{six orders of magnitude}. Remarkably, despite operating under much stricter power and area constraints, Pinball also reduces logical error rates by 32.58$\times$ and 5$\times$, respectively, compared to the state-of-the-art RT predecoder and an RT ensemble configuration. By increasing cryogenic coverage, we also reduce syndrome bandwidth up to 3780.72$\times$. Through co-design with 4\,K-characterized 22\,nm FDSOI technology, we achieve a peak power consumption under 0.56\,mW. Voltage/frequency scaling and body biasing enable 22.2$\times$ lower typical power consumption, yielding up to 67.4$\times$ total energy savings. Assuming a 4\,K power budget of 1.5\,W, our predecoder can support up to 2,668 logical qubits at $d=21$.
\end{abstract}

\section{Introduction} \label{sec:intro}
While quantum computing offers significant advantages for problems in domains like chemistry and optimization, solving practical instances will require $10^{5}-10^6$ qubits with error rates near $10^{-15}$ \cite{beverland2022assessing}. Due to environmental noise, however, current systems are limited to just $10^2-10^3$ qubits with errors rates of $10^{-2}-10^{-3}$. To bridge this gap between present-day devices and future application requirements, quantum error correction (QEC) uses redundancy to encode low-error \textit{logical} qubits from many error-prone \textit{physical} qubits. Errors are detected through repeated measurements, producing a binary \textit{error syndrome} used by a classical decoder to determine corrections. If physical error rates stay below a threshold, logical error rates can drop exponentially. The surface code \cite{fowler2012surface} is a leading QEC code candidate, especially in the intermediate-term, due to its planar layout, local connectivity, and high error threshold.

Despite its advantages, the surface code is challenging to implement practically. For instance, the volume of syndrome data scales cubically with the code's distance, $d$, and some gates in quantum circuits are conditioned on the results of decoding, introducing strict real-time latency requirements. Furthermore, many prominent qubit modalities are stored cryogenically inside dilution refrigerators \cite{iwai2023cryogenic, van2019electronic}, creating significant spatial separation with the decoder at room-temperature (RT). This requires high-bandwidth, cross-temperature interconnects via long, lossy coaxial cables, which is unsustainable due to: (1) limited microwave I/O, (2) heat flux from RT connections that degrades coherence, and (3) long, variable feedback delays that hinder synchronization.

\begin{figure}
    \centering
    \includegraphics[width=\columnwidth]{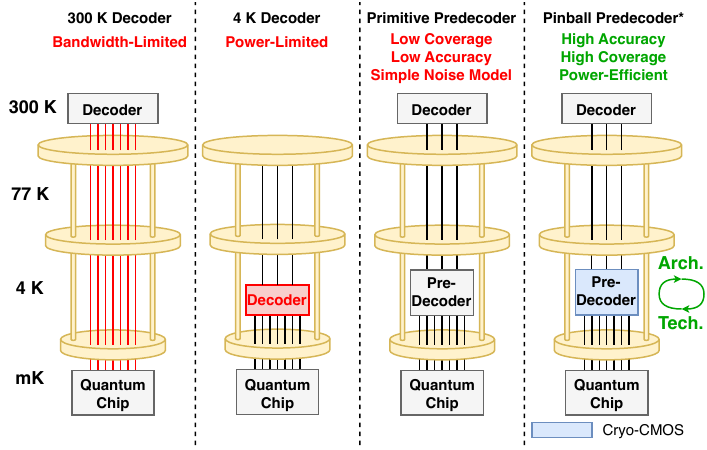}
    \caption{The classical hardware landscape for QEC decoding. Relative to prior work, our design (rightmost) (1) improves predecoding performance in realistic settings by analyzing error sources and propagation at the circuit-level, and (2) co-optimizes architecture with technology to achieve this higher performance at lower overheads.}
    \label{fig:intro}
    \vspace{-0.5cm}
\end{figure}

To address control and readout bandwidth, academia and industry have explored co-locating electronics alongside qubits in the cryogenic domain, typically at 4\,K \cite{bardin201929, bardin2019design, chakraborty2022cryo, underwood2024using, van2020scalable, park2021fully, frank2023low, yoo202334}. However, the 4\,K cooling capacity is very low ($\sim$1.5\,W \cite{krinner2019engineering}), and if cryogenic hardware exceeds this budget, the added thermal noise harms qubit fidelity. While optimization is still needed to reduce control and readout power, substantial advances have been made in recent years to lower overheads (e.g., 23 mW/qubit control power \cite{frank2022cryo} down to $<$4 mW/qubit \cite{yoo202334}).

\textit{Nonetheless, even with additional reductions, QEC syndrome transmission power will remain a prominent concern if left unaddressed.} Similar to control and readout, cryogenic decoding has been explored to reduce bandwidth requirements. However, due to 4\,K power constraints, fully cryogenic decoders \cite{holmes2020nisq+, ueno2021qecool, ueno2022qulatis} suffer from insufficient accuracy and are unable to support many high-distance logical qubits. Thus, significant architecture and technology innovations (Fig. \ref{fig:intro}) are needed to realize cryo-compatible decoding strategies.

At the architecture level, \textbf{cryogenic predecoding} \cite{ravi2023better} has been proposed to apply local decoding heuristics instead of complex, global algorithms. Predecoding can maintain high accuracy because, even at current error rates, long error chains are rare, and most syndromes are sparse and local. Cryogenic predecoders use low-power logic to decode simple syndromes and flag rare, complex ones for further processing. If cryogenic predecoders work well, only complex syndromes are sent to a more power-intensive RT decoder, keeping interconnect bandwidth between 4\,K and RT low without sacrificing accuracy. 

While cryogenic predecoding holds great promise, prior work~\cite{ravi2023better} has tailored to simplistic noise models that fail to account for many sources of errors and how they propagate in real QEC implementations. When evaluated under more realistic noise, we find that prior work's utility is significantly dampened in terms of coverage and predecoding accuracy, limiting the predecoder's ability to reduce syndrome bandwidth and achieve sufficiently low logical error rates (see Sec. \ref{sec:predecoding}).

At the technology level, \textbf{cryogenic CMOS (cryo-CMOS)} and single-flux quantum (SFQ) logic have emerged as leading candidates for low-power, cryogenic electronics. However, today's SFQ has many scaling disadvantages such as significant area consumption \cite{holmes2020nisq+}, complex pulse-based physical design \cite{choi2024supercore}, less mature EDA toolchains, and challenging integration with ubiquitous CMOS technologies. By contrast, cryo-CMOS logic benefits from extremely high area density, decades of research in design optimization techniques, and a mature EDA toolchain. These advantages have spurred significant academic \cite{kang202334, schmidt202513, van2020scalable} and industrial \cite{chakraborty2022cryo, underwood2024using, frank2023low, bardin201929, bardin2019design, yoo202334} investment in cryo-CMOS qubit interfacing systems. 

Despite cryo-CMOS's advantages, prior cryogenic decoding works \cite{ravi2023better, holmes2020nisq+, ueno2021qecool, ueno2022qulatis} have almost exclusively used SFQ, limiting design characterization and precluding the essential architecture-technology co-design needed for practical solutions. On the one hand, hardware characterization enables the detailed analyses required to accurately assess design scalability, and on the other, cross-layer optimizations can yield scalability surpassing that achievable by optimizations made in a single layer. These implications are critical in the context of cryogenic predecoding, where designs cannot be evaluated based only on raw performance metrics (e.g., bandwidth reduction), but by how much performance they can achieve while still operating within the power and area constraints of the dilution refrigerator.

For these reasons, this paper presents a cross-layer design, from algorithm to cryo-CMOS implementation in 22\,nm Fully Depleted Silicon on Insulator (FDSOI), of \textbf{Pinball}, a cryogenic predecoder for surface code QEC capable of handling realistic, circuit-level noise. By analyzing both how errors occur on and propagate across qubits in syndrome measurement circuits, as well as how errors correlate syndromes together, we derive a small set of simple algorithmic predecoding primitives capable of correcting them. Further, we identify race conditions between conflicting primitives that motivate the algorithm's translation into a pipelined architecture, using error frequency analysis to determine a good ordering of pipeline stages. Finally, by co-designing with cryo-CMOS FDSOI technology, we exploit latency requirement imbalances in predecoding to markedly reduce power and energy consumption through voltage/frequency scaling and forward body biasing.

To summarize, the contributions of this paper are:
\begin{enumerate}
    \item \textbf{Circuit-Level Noise Analysis}: Detailed analysis of error propagation in surface code QEC circuits, categorizing them based on how they correlate syndromes across space and time. In conjunction with error frequency analysis, these enable predecoding improvements over prior work.

    \item \textbf{Pinball Predecoding Algorithm and Architecture}: Development of a novel predecoding algorithm, Pinball, and its pipelined architecture, capable of surface code QEC predecoding under realistic, circuit-level noise. At near-term (long-term) error rates, Pinball reduces syndrome transmission by up to 5.65$\times$ (34.72$\times$) and increases predecoding accuracy by up to 4.23$\times$ (1.46$\times$) relative to the state-of-the art cryogenic predecoder, achieving total syndrome bandwidth reduction up to 3780.72$\times$ and reducing logical error rate (LER) up to nearly six orders of magnitude. Remarkably, Pinball even reduces LER by 32.58$\times$ and 5$\times$ relative to the state-of-the-art RT predecoder and ensemble configurations, respectively.
    
    \item \textbf{Architecture-Technology Co-Optimization}: By identifying significant latency requirement imbalances in the per-round predecoding process, we co-design with cryo-CMOS capabilities, leveraging voltage/frequency scaling and forward body biasing for lower supply voltage and reducing power by up to $22.2\times$. Factoring in cryo-to-RT transmission, we can achieve total energy savings up to $37.05\times$. Considering additional overheads from a real-world QEC interface \cite{riverlane2024qeci}, this number rises to $67.4\times$.
    
    \item \textbf{Cryo-CMOS Implementation and Evaluation}: Using a 22\,nm FDSOI technology designed for 4\,K, we perform synthesis, placement, and routing of Pinball with power, performance, and area evaluation. Under a 1.5\,W budget for cryogenic predecoding, our design supports 37,313 and 2,668 logical qubits at code distances of 3 and 21 respectively, at a physical error rate of $10^{-3}$.
\end{enumerate}
\section{Background and Motivation} \label{sec:background}

\subsection{Quantum Error Correction} \label{sec:qec}
Similar to classical error correction, quantum error correction (QEC) uses redundancy to encode a logical qubit using many physical qubits with high \textit{physical error rate}, $p$. However, QEC faces two unique challenges: measuring a qubit collapses its state, and qubits suffer both bit flip ($X$) and phase flip ($Z$) errors, unlike classical bits.

\begin{figure}
    \begin{minipage}{0.48\columnwidth}
        \begin{subfigure}{\textwidth}
            \includegraphics[width=\textwidth]{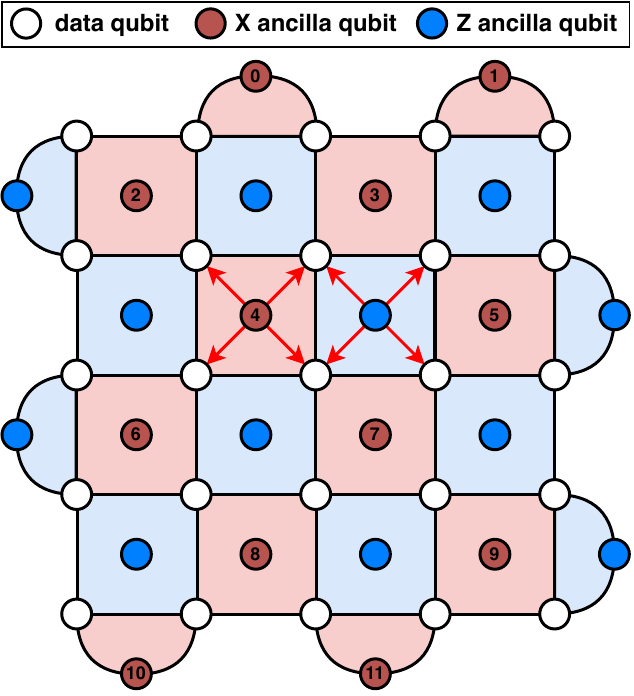}
            \caption{}
            \label{fig:surface-code}
        \end{subfigure}
    \end{minipage}
    \hfill
    \begin{minipage}{0.45\columnwidth}
        \centering
        \begin{subfigure}{\linewidth}
            \centering
            \includegraphics[width=\textwidth]{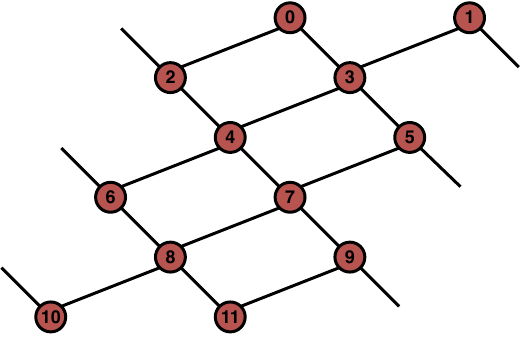}
            \caption{}
            \label{fig:decoding-graph}
        \end{subfigure}\\[1ex]
        \begin{subfigure}{0.4\linewidth}
            \centering
            \includegraphics[width=\textwidth]{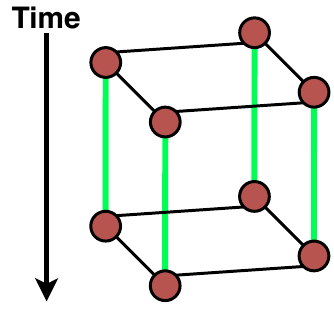}
            \caption{}
            \label{fig:phenom-unit-cell}
        \end{subfigure}
        \hspace{10px}
        \begin{subfigure}{0.4\linewidth}
            \centering
            \includegraphics[width=\textwidth]{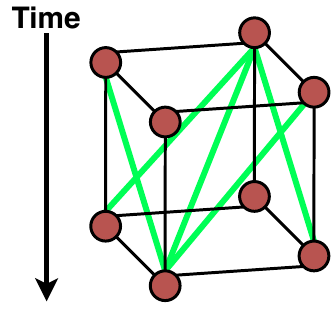}
            \caption{}
            \label{fig:ckt-level-unit-cell}
        \end{subfigure}
    \end{minipage}
    \caption{(a) $d=5$ rotated surface code lattice. (b) Corresponding decoding graph for $Z$ errors in one QEC round. (c, d) Unit cells for the $Z$-error decoding graph under (c) phenomenological noise and (d) circuit-level noise.}
    \label{fig:surface-code-background}
\end{figure}

To address these, various QEC codes have been developed, with the surface code being a leading candidate due to its planar layout and local connectivity. Implemented on a 2D lattice (Fig. \ref{fig:surface-code}), the rotated surface code uses $2d^2 - 1$ physical qubits (data qubits and $X$/$Z$ ancilla qubits) where $d$ is the \textit{code distance}, or the minimum length of a \textit{logical error}.

\begin{figure}
    \centering
    \begin{subfigure}{0.5\columnwidth}
        \includegraphics[width=\textwidth]{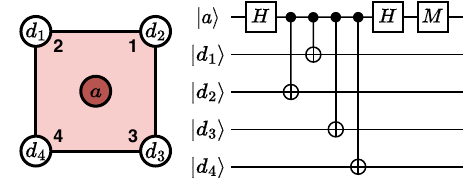}
        \label{fig:x-stab}
        \vspace{-10pt}
        \caption{}
    \end{subfigure}
    \begin{subfigure}{0.45\columnwidth}
        \includegraphics[width=\textwidth]{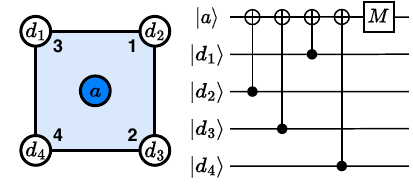}
        \label{fig:z-stab}
        \vspace{-10pt}
        \caption{}
    \end{subfigure}
    \caption{Syndrome measurement circuits for detecting (a) $Z$ errors and (b) $X$ errors. $X$/$Z$ ancilla patches are annotated with numbers specifying the order of CNOTs in their circuits.}
    \label{fig:qec-circuits}
\end{figure}

Ancilla qubits are entangled with neighboring data qubits (red arrows) and measured to detect $Z$ and $X$ errors. The circuits implementing these interactions are shown in Fig. \ref{fig:qec-circuits}. They exploit the propagation of Pauli terms through CNOTs, namely that \textit{Pauli-$Z$ terms propagate through a CNOT from the target qubit to the control qubit} while \textit{Pauli-$X$ terms propagate from the control qubit to the target qubit}, in order to transmit error information from data qubits to ancilla qubits.

At the end of each circuit, the ancilla qubits are measured, and the parity between consecutive measurement outcomes forms a binary \textit{syndrome} which a decoder uses to infer corrections. Each ancilla reports only error parity (0 for even, 1 for odd) so different error patterns can yield identical syndromes, complicating decoding. In CSS codes, $X$ and $Z$ syndromes can be decoded separately.

\subsection{Noise Models \& the Decoding Graph} \label{sec:noise-models}
QEC decoding complexity and accuracy greatly depend on the degree to which physical qubit noise and its corresponding error sources are accounted for. This complexity is often reflected in the noise model used to analyze and design the classical decoder. 
For example, the simplest, lowest-accuracy decoders assume errors are stochastically injected onto each data qubit with a probability, $p$; this process is captured by the \textit{code capacity noise model}. In reality, ancilla measurement itself is error-prone, requiring decoders to process multiple (typically $d$) syndrome measurement rounds under the more accurate \textit{phenomenological noise model}. This description is still lacking, since single-qubit, two-qubit, and reset operations in QEC circuits also introduce errors; these are incorporated into the \textit{circuit-level noise model} used by the highest-accuracy decoders. While error interactions become increasingly complex to decode with each model, accounting for all of them is critical to enable the high-accuracy QEC decoding required for practically useful, fault-tolerant execution.

The decoding problem can be represented by a weighted graph with three dimensions: two spatial (the surface code lattice) and one temporal (multiple measurement rounds). Fig. \ref{fig:decoding-graph} shows the $Z$-error decoding graph for one measurement round, corresponding to the code capacity noise model. Vertices represent syndromes, and edges represent error sources that flip the syndrome values at their endpoints. Optionally, edge weights can be used to reflect error probabilities. Syndromes with an odd number of adjacent errors are termed \textit{active} and are paired by the decoder to infer corrections on data qubits.

Fig. \ref{fig:phenom-unit-cell} shows a single unit cell of the $Z$-error decoding graph for the phenomenological noise model. In addition to the horizontal edges within a round (\textit{space-like errors}), vertical edges (green) correspond to measurement errors (\textit{time-like errors}). In the unit cell for the circuit-level noise model (Fig. \ref{fig:ckt-level-unit-cell}), additional diagonal edges (green) with both space and time components (\textit{spacetime-like errors}) account for errors propagating through CNOT gates in the syndrome measurement circuits. Critically, the decoding algorithm requires a mapping from each error edge to the correspondingly affected data qubit(s), necessitating an analysis of how errors emerge and propagate throughout the syndrome measurement circuit.  

\subsection{Decoding Algorithms} \label{sec:decoding-algorithms}
To address the complexity of surface code decoding, several algorithms have been developed to accurately identify and correct errors. Critically, these algorithms must operate within strict real-time constraints. One reason is that non-Clifford gates (e.g., $T$ gates) can fail injection, and if so, require corrective $S$ gates dependent on the decoding history \cite{fowler2012surface}. Moreover, decoding must be completed at least as quickly as the rate of qubit measurement to avoid an exponentially growing backlog of error data \cite{terhal2015quantum}. For superconducting qubits, this latency can be as low as 1\,$\mu$s \cite{skoric2023parallel}.

The Blossom algorithm \cite{edmonds1965paths} based minimum-weight perfect matching (MWPM) decoder offers high accuracy, but with $O(d^5)$ time complexity. The Union-Find decoder \cite{delfosse2021almost} trades off accuracy for improved latency ($O(d^3)$ complexity). While effective, these and related algorithms \cite{higgott2025sparse, wu2023fusion} impose high resource and latency costs. At RT, these costs are tolerable, but cryogenic constraints, especially the low power budgets ($\sim$1.5\,W at 4\,K \cite{krinner2019engineering}), make such decoders impractical to co-locate with cryogenic quantum hardware. 

\subsection{4 K-to-RT Bandwidth Constraints} \label{sec:bandwidth-constraints}
\begin{figure}
    \centering
    \includegraphics[width=\columnwidth]{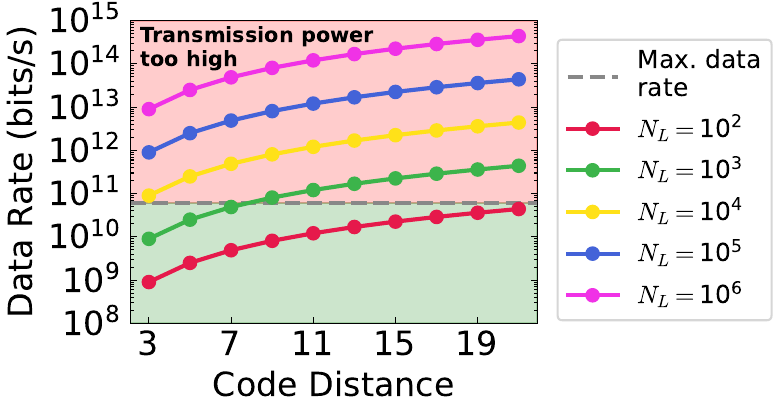}
    \caption{Data rates needed to support $N_L$ logical qubits at various code distances. Horizontal line: the data rate at which transmission power exceeds the 4\,K power budget.}
    \label{fig:bandwidth-scaling}
    \vspace{-0.25cm}
\end{figure}

In addition to power and latency constraints, an important design constraint is the volume of data entering and exiting the dilution refrigerator between the 4 K and RT stages. This volume includes qubit control instructions as well as syndrome and correction data. While the repetitive structure of logical operations \cite{tan2024sat} will likely allow minimizing instruction bandwidth using a cryogenic quantum control processor \cite{tannu2017taming}, streaming surface code syndrome data to RT will continue to dominate, especially since (1) its volume scales as $O(d^2)$ per round per logical qubit (see Fig. \ref{fig:bandwidth-scaling}) and (2) $d$ QEC rounds are required per logical operation \cite{litinski2019game}.

Using a state-of-the-art wireline transmitter dissipating 0.92 pJ/b \cite{cusmai20247}, we estimate the interconnect power as a function of the required syndrome bandwidth. We find that, given 1.5 W of cooling power at 4\,K \cite{krinner2019engineering}, the transmission power exceeds the entire budget at 60.98 Gb/s, only enabling support of 1000 logical qubits for code distances up to 7. Moreover, since every cable entering the 4\,K stage from RT introduces additional passive heat load~\cite{krinner2019engineering, wang2025wireless}, the number of links must be minimized. Clearly, scaling to the large code distances required for fault-tolerant quantum computing requires innovative strategies to mitigate this bandwidth bottleneck.

\subsection{Predecoding Algorithms} \label{sec:predecoding}
Fortunately, most decoding scenarios involve sparse active syndromes~\cite{ravi2023better, delfosse2020hierarchical}, as long error chains are exponentially unlikely. This has motivated the development of QEC predecoding \cite{ravi2023better, smith2023local, alavisamani2024promatch} which uses much simpler logic than full-fledged decoders to pre-process these sparse syndromes. We discuss two types of previously studied predecoding: RT predecoding and cryogenic predecoding.

\underline{RT predecoding} \cite{smith2023local, alavisamani2024promatch} typically seeks to \textit{minimize decoding latency} by sparsifying syndromes before they are sent to the full-fledged RT decoder. Specifically, RT predecoders identify short correction chains to match local pairs of syndrome bits, remove them, and pass the reduced syndrome on to the decoder. By giving the decoder a sparser, simpler decoding graph, its latency remains below the 1\,$\mu$s budget for higher code distances than would be possible without predecoding. A state-of-the-art RT predecoder is Promatch \cite{alavisamani2024promatch}.

On the other hand, \underline{cryogenic predecoding} \cite{ravi2023better} seeks to \textit{minimize 4\,K-to-RT syndrome bandwidth} by decoding a large fraction of syndromes within the 4\,K temperature stage. Cryogenic predecoders use lightweight decoding heuristics to handle common, short (typically length-1) error chains within the 4\,K power and area constraints. To maintain the extremely low logical error rates ($10^{-15}$) required for applications, cryogenic predecoders must also detect longer error chains, offloading their complex syndromes to RT for full-fledged decoding. Since complex scenarios are very rare, 4\,K-to-RT syndrome bandwidth can be significantly lowered. The state-of-the-art cryogenic predecoder is Clique \cite{ravi2023better}.

Since they solve different problems, RT and cryogenic predecoders are orthogonal. Ideally, future quantum computing systems will combine both, using the former to minimize 4\,K-to-RT bandwidth and the latter to minimize decoding latency when complex syndromes are propagated to RT. While we focus on cryogenic predecoding, we include comparisons of our predecoder with both Promatch and Clique in Sec. \ref{sec:evaluation}.

\begin{figure}
    \centering
    \includegraphics[width=\columnwidth]{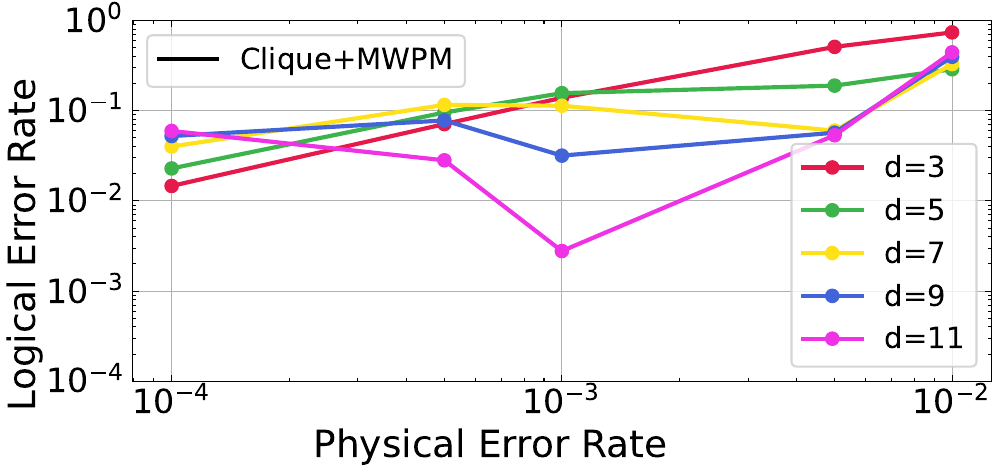}
    \vspace{-0.5cm}
    \caption{LER achieved under circuit-level noise by Clique \cite{ravi2023better}.}
    \label{fig:clique_ler}
    \vspace{-0.25cm}
\end{figure}

\noindent
\textbf{Predecoding Under Realistic Noise:} We now motivate our reasoning for targeting cryogenic predecoding by analyzing how the current state-of-the-art, Clique \cite{ravi2023better}, performs under noise present in real quantum computing systems. As a simplification, Clique considers only space-like and time-like errors, ignoring syndrome pairs due to spacetime-like errors. Since its evaluation in \cite{ravi2023better} used only phenomenological noise, Clique still maintained high accuracy while achieving high bandwidth savings. To understand Clique's benefits for real-world systems, we evaluated it under realistic, circuit-level noise (described in Sec. \ref{sec:methodology}). Fig. \ref{fig:clique_ler} shows that Clique is unable to provide the exponential logical error rate suppression needed by applications (discussed further in Sections \ref{cmp_clique} and \ref{sec:cryo-perf}). Thus, to the best of our knowledge, there are no cryogenic predecoders which can provide both low enough syndrome bandwidth and low enough logical error rates to achieve a practical, scalable quantum computing system. Clearly, further innovations are needed in cryogenic predecoding, and this is the crux of our motivation to design Pinball.

\subsection{Cryogenic Technology}
To meet the stringent design constraints of cryogenic environments, both cryo-CMOS and SFQ logic have emerged as promising technologies.
Recent work has explored SFQ-based accelerators for domain-specific applications~\cite{holmes2020nisq+,jokar2022digiq,ishida2020supernpu,kashima202164}, but SFQ design presents several challenges. First, scalability is difficult due to SFQ’s inherently low area density: standard SFQ logic cells (e.g., AND, OR, XOR) are typically ${\sim}$$10^5$–$10^6$${\times}$ larger than their CMOS counterparts~\cite{holmes2020nisq+}. Second, SFQ’s pulse-based logic requires each cell to be clocked separately, demanding deep pipelining and non-standard architectural techniques~\cite{choi2024supercore} that complicate physical design. Third, the SFQ design ecosystem suffers from a lack of mature EDA tools, limiting design productivity and optimization \cite{huang2022survey}. Finally, integrating SFQ circuits with conventional CMOS technologies, whether at cryogenic temperatures or RT, introduces considerable system-level complexity.

In contrast, cryo-CMOS offers several compelling advantages: it supports high integration density, benefits from decades of design optimization research, and leverages well-developed EDA tools. Moreover, major quantum computing initiatives, including those at IBM~\cite{chakraborty2022cryo, underwood2024using, frank2023low} and Google~\cite{bardin201929, bardin2019design, yoo202334}, have adopted cryo-CMOS for control hardware, reinforcing its ecosystem maturity. Thus, using cryo-CMOS for cryogenic QEC hardware simplifies system integration and aligns with current industry practices.

Therefore, we select cryo-CMOS, specifically a 22\,nm FDSOI technology re-characterized at 4\,K, for Pinball. While cryo-CMOS is sometimes considered less energy-efficient than SFQ, we demonstrate that through careful architecture-technology co-design, Pinball achieves better energy consumption than SFQ designs. Note that CMOS behavior at 4\,K differs significantly from RT in several key aspects:
\begin{itemize}[leftmargin=*]
    \item \textbf{Increased threshold voltage} reduces voltage scaling headroom but can be mitigated by forward body biasing \cite{overwater2023cryogenic}.
    
    \item \textbf{Enhanced carrier mobility} improves transistor speed and performance.

    \item \textbf{Technology-dependent current gain}: \cite{parihar2023cryogenic} reports +1.5\% in 40\,nm, -3.8\% in 14\,nm FinFET, and we observe -10.5\% in the 22\,nm FDSOI technology used in this work.

    \item \textbf{Reduced subthreshold slope} lowers leakage and enables faster switching which benefits low-power operation \cite{parihar2023cryogenic}.

    \item \textbf{Increased mismatch} \cite{li202513} requires conservative design with extensive calibration to maintain reliability.
\end{itemize}

These cryogenic effects collectively influence digital logic performance. In particular, enhanced mobility and lower resistance compensate for increased threshold voltage, resulting in faster operation at 4\,K \cite{fakkel2023cryo} with lower leakage. Moreover, reduced subthreshold slope and body biasing enable reliable subthreshold operation for ultra-low-power design. 
\section{Pinball Predecoding Algorithm} \label{sec:pinball-alg}
\begin{figure}
    \centering
    \includegraphics[width=\columnwidth]{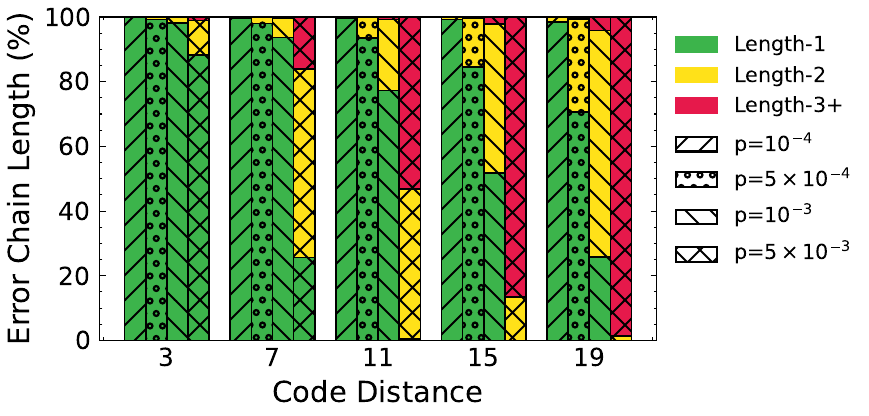}
    \caption{Distribution of maximum-length error chains in the SI1000 noise model (see Sec. \ref{sec:methodology}).}
    \label{fig:correlated_distribution}
\end{figure}

We begin by describing Pinball's lightweight predecoding algorithm which is responsible for decoding a single logical qubit. As with previous cryogenic predecoding work \cite{ravi2023better}, we target predecoding of error chains spanning a single edge in the decoding graph, since they constitute a significant fraction of observed errors and are relatively easy to decode and correct. In Fig. \ref{fig:correlated_distribution}, we verify this claim by simulating $10^5$ blocks of $d$-round syndromes across many physical error rates and code distances, recording the longest-observed error chain per block. For almost all code distances and error rates, the longest error chains have length 1 at least 20\% (often much higher) of the time. Underlying the Pinball algorithm is the observation first made in \cite{delfosse2020hierarchical} and also leveraged in \cite{ravi2023better, alavisamani2024promatch, smith2023local} that, \textit{if syndromes adjacent to each other in the decoding graph are both active, then there is a high likelihood that the edge between them has been flipped by an error}.

Motivated by this observation, we derive a basic predecoding primitive with the following logic: for a given pair of neighboring syndromes, (1) check if they are both active, and if so, (2) assign corrections to the data qubit(s) corresponding to the edge between them. We can then pattern this predecoding primitive over every edge in the decoding graph to assign a full set of corrections for length-1 errors. While the logic itself is very simple, achieving predecoding coverage of the entire decoding graph at once requires $O(d^3)$ such primitives. 

To improve cryogenic scalability, Pinball operates in a streaming fashion, predecoding only on pairs of successive syndrome rounds, $S_{i-1}$ and $S_i$, two at a time, as they become available from the readout circuitry. Operating on this smaller \textit{predecoding subgraph} reduces hardware overheads to $O(d^2)$. Notably, this simplification does not sacrifice predecoding accuracy: since ancilla qubits are reset following every round of syndrome measurement, all length-1 error chains span at most two syndrome rounds. 

As introduced in Sec. \ref{sec:noise-models}, the circuit-level noise model requires a predecoding subgraph composed of three types of edges: space-like, time-like, and spacetime-like. From the predecoding subgraph, we can derive which syndrome pairs must be covered by predecoding primitives to ensure full coverage. The only remaining step is to determine which qubit(s) to correct for each edge between covered syndrome pairs. To do so, in the following sections, we analyze error sources and their propagation in the syndrome measurement circuits. Although we restrict our analysis to $Z$ errors, it is applicable for any Pauli-type error, since $X$ errors are symmetric and $Y$ errors can be decomposed into $Z$ and $X$  components.

\subsection{Space-like Errors}
\begin{figure}
    \centering
    \includegraphics[width=\columnwidth]{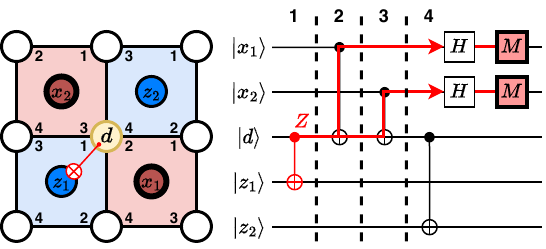}
    \caption{Formation of a space-like error.}
    \label{fig:spacelike-error}
\end{figure}

Space-like errors correlate two syndromes within the same syndrome measurement round. Fig. \ref{fig:spacelike-error} shows one example of error generation and propagation causing such a correlation. In the first timestep, the CNOT gate between qubits $d$ and $z_1$ induces a $Z$ error on qubit $d$. In timesteps 2 and 3, subsequent CNOT interactions with qubits $x_1$ and $x_2$ propagate the $Z$ error to them, leading to error detections at the end of their circuits. From this example, it is clear that, for space-like errors, the data qubit requiring correction is the one shared between the two active syndromes in the surface code lattice. The only exceptions are at the edge of the code lattice, where errors on data qubits are detected by only a single ancilla. 

\begin{figure}[ht!]
    \centering
    \begin{subfigure}{0.49\columnwidth}
        \includegraphics[width=\textwidth]{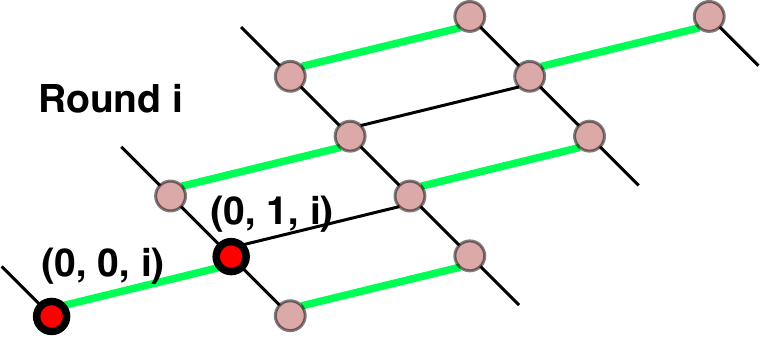}
        \caption{}
        \vspace{0.25cm}
    \end{subfigure}
    \begin{subfigure}{0.49\columnwidth}
        \includegraphics[width=\textwidth]{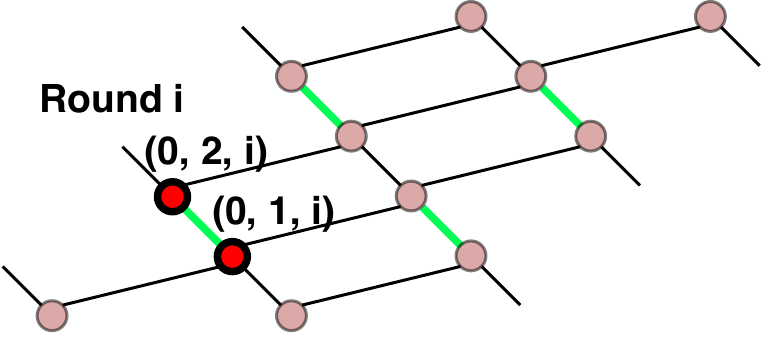}
        \caption{}
        \vspace{0.25cm}
    \end{subfigure}
    
    \begin{subfigure}{0.49\columnwidth}
        \includegraphics[width=\textwidth]{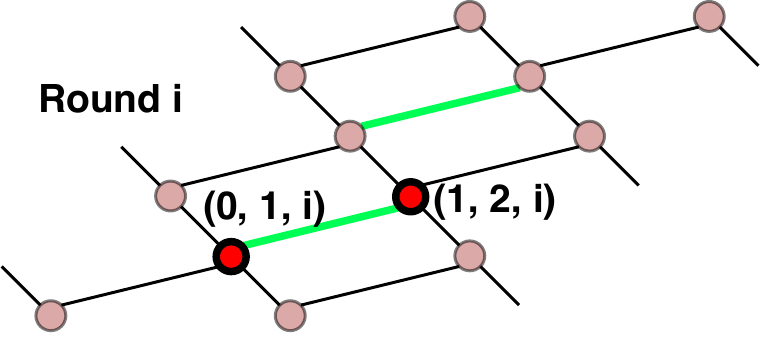}
        \caption{}
        \vspace{0.25cm}
    \end{subfigure}
    \begin{subfigure}{0.49\columnwidth}
        \includegraphics[width=\textwidth]{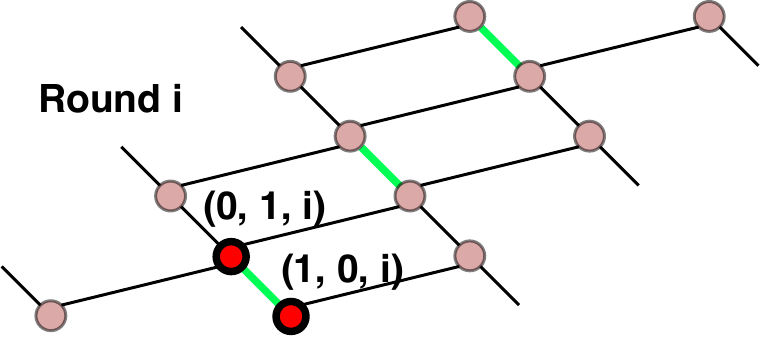}
        \caption{}
        \vspace{0.25cm}
    \end{subfigure}
    \begin{subfigure}{0.49\columnwidth}
        \includegraphics[width=\textwidth]{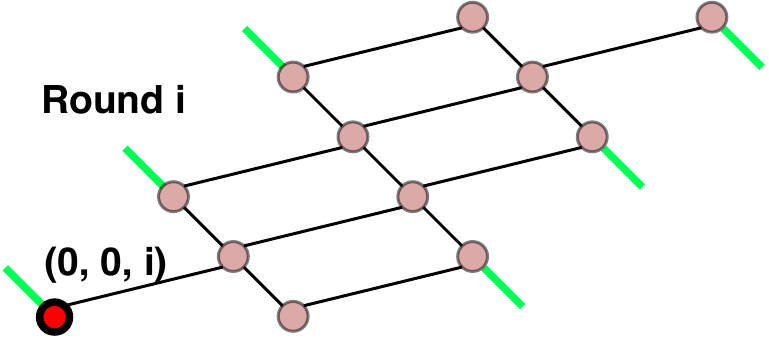}
        \caption{}
    \end{subfigure}
    \caption{Full coverage for space-like errors in the predecoding subgraph. Green edges indicate covered errors, and 3D coordinates $(x,y,t)$ specify relative spatial ($x,y$) and temporal ($t$) positioning of syndromes. Space-like errors within the bulk of the subgraph are divided into four non-conflicting groups mapped to pipeline stages: (a) \textbf{B(1)}, (b) \textbf{B(2)}, (c) \textbf{B(3)}, and (d) \textbf{B(4)}. (e) Space-like errors at the edges of the subgraph are mapped to a separate pipeline stage, \textbf{E}.}
    \label{fig:space-like-coverage}
\end{figure}

For Pinball to obtain full coverage of length-1 space-like errors, we require predecoding primitives for all pairs of neighboring ancilla qubits in the lattice, as well as for each ancilla qubit at the lattice edges (Fig. \ref{fig:space-like-coverage}).

\subsection{Time-like Errors}
\begin{figure}[ht!]
    \centering
    \includegraphics[width=\columnwidth]{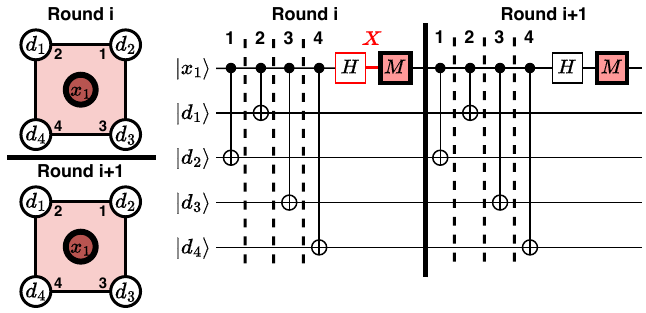}
    \caption{Formation of a time-like error.}
    \label{fig:timelike-error}
\end{figure}

Time-like errors correlate the same syndrome across different syndrome measurement rounds. In Fig. \ref{fig:timelike-error}, we show an example of how a single $X$ error on an $X$ ancilla qubit due to the Hadamard gate preceding its measurement operation flips syndromes in two consecutive rounds, $i$ and $i+1$. Recall that the vertices in the predecoding subgraph store the parity of measurements between rounds rather than the raw measurement values from each round. Thus, assuming an error-free round $i-1$, the $X$ error in round $i$ flips the first measurement outcome from 0 to 1 relative to the previous measurement, inducing an active syndrome in round $i$. Similarly, assuming an error-free round $i+1$, the second measurement operation flips the outcome from 1 back to 0 relative to round $i$, inducing a second active syndrome in round $i+1$. 

\begin{figure}[ht!]
    \centering
    \includegraphics[width=0.7\columnwidth]{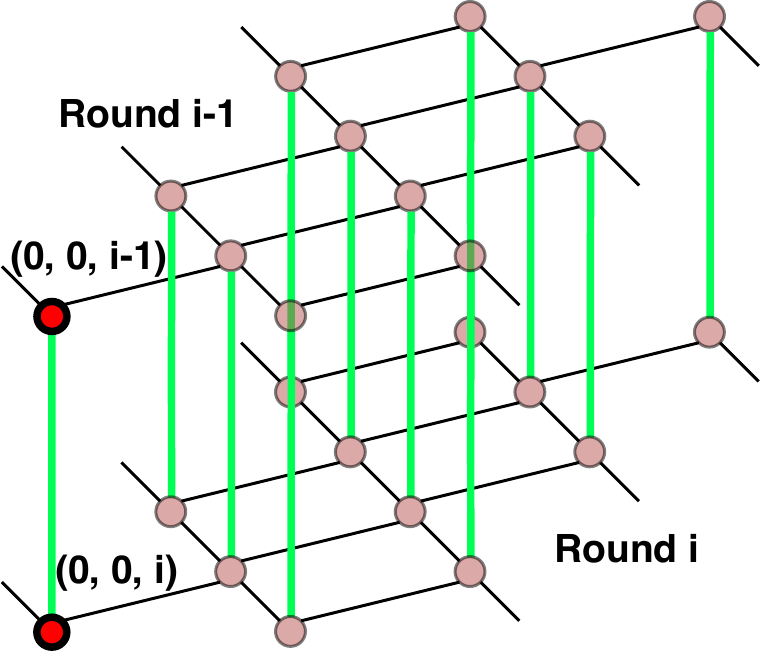}
    \caption{Full coverage for time-like errors in the predecoding subgraph which are mapped to a single pipeline stage, \textbf{M}.}
    \label{fig:time-like-coverage}
\end{figure}

To achieve full coverage of time-like length-1 errors, we require predecoding primitives that consider pairs of the same syndrome across consecutive rounds (Fig. \ref{fig:time-like-coverage}). Unlike space-like errors, time-like errors do not require correction assignment since the errors are associated with ancilla qubits which are reset before each measurement round.

\subsection{Spacetime-like Errors}
We now introduce spacetime-like errors, which arise when considering single-qubit and two-qubit gate errors and how they propagate through the syndrome measurement circuit. Spacetime-like errors correlate syndromes separated both in time (across rounds), and in space (different locations within the surface code lattice). We differentiate between those which propagate to a single data qubit and those which propagate to two data qubits (i.e., hook errors \cite{dennis2002topological}).

\begin{figure}
    \centering
    \includegraphics[width=\columnwidth]{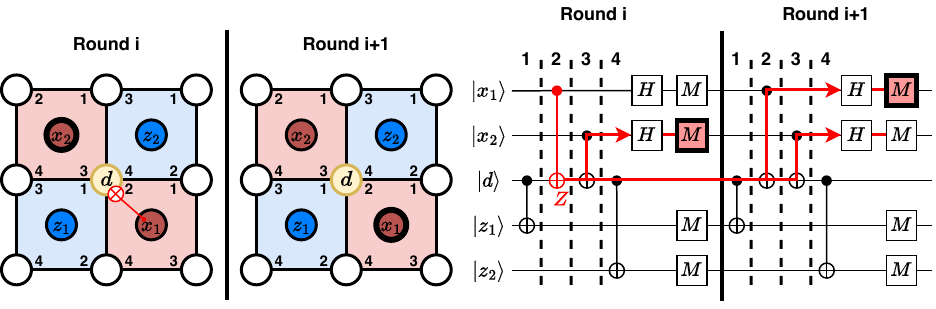}
    \caption{Formation of a spacetime-like error.}
    \label{fig:spacetime-error}
\end{figure}

An example single-qubit spacetime-like error is shown in Fig. \ref{fig:spacetime-error}. During syndrome measurement round $i$, the CNOT gate between qubits $x_1$ and $d$ induces a $Z$ error on qubit $d$. Since this error only manifests \textit{after} the CNOT, it is not immediately propagated to qubit $x_1$ in timestep 2. In timestep 3, the CNOT between qubits $x_2$ and $d$ propagates the error to qubit $x_2$ which detects it at the end of the syndrome measurement round. In the absence of correction (corrections are typically applied only after every $d$ rounds of syndrome measurement), the $Z$ error on qubit $d$ persists into round $i+1$, where it propagates to both qubit $x_1$ and qubit $x_2$ in timesteps 2 and 3, respectively. Only qubit $x_1$ produces an active syndrome, since qubit $x_2$'s error parity has not changed from round $i$.

\begin{figure}
    \centering
    \includegraphics[width=0.6\columnwidth]{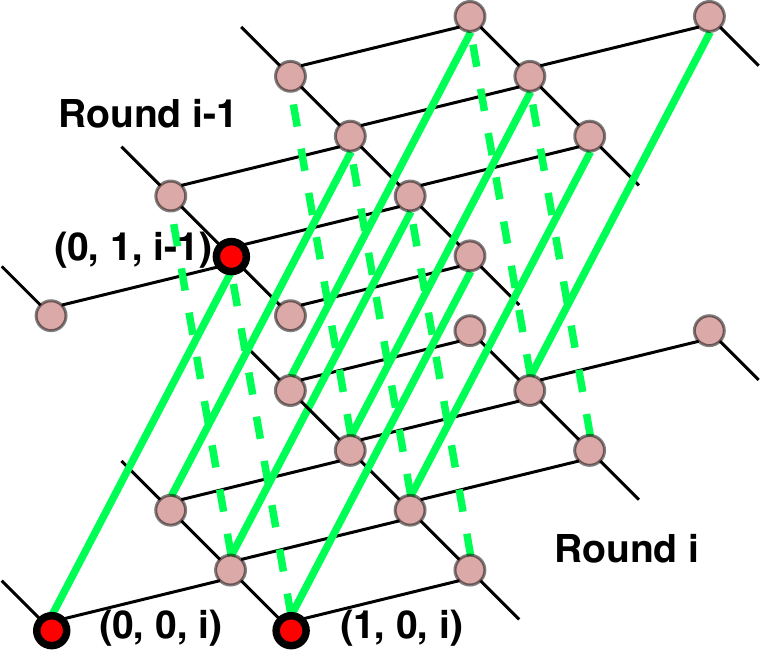}
    \caption{Full coverage for single-qubit, spacetime-like errors in the predecoding subgraph. These errors are divided into two non-conflicting groups mapped to pipeline stages \textbf{ST(1)} (solid edges) and \textbf{ST(2)} (dashed edges).}
    \label{fig:spacetime-like-coverage}
\end{figure}

To achieve full coverage of single-qubit, spacetime-like length-1 errors, we require predecoding primitives covering all pairs of syndromes across consecutive rounds which are adjacent to each other in the surface code lattice (Fig. \ref{fig:spacetime-like-coverage}). If both syndromes in a pair are active, the predecoding primitive assigns a correction to the shared data qubit between the ancilla qubits in the lattice.

\begin{figure}
    \centering
    \includegraphics[width=0.9\columnwidth]{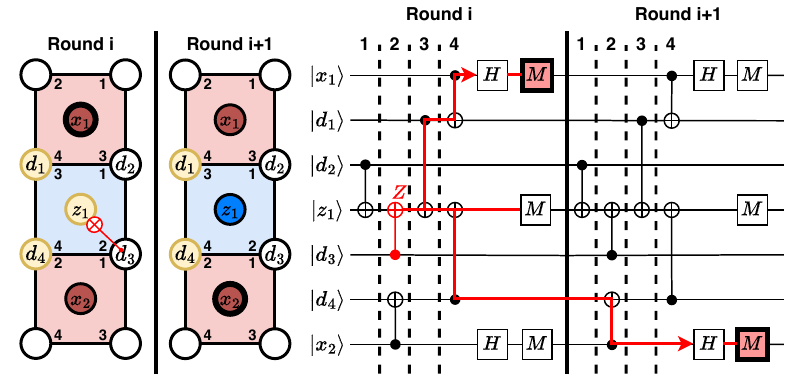}
    \caption{Formation of a hook spacetime-like error.}
    \label{fig:hook-error}
\end{figure}

Finally, we consider hook spacetime-like errors which are shown in Fig. \ref{fig:hook-error}. Unlike all previously analyzed errors, hook errors arise from a unique error source in the syndrome measurement circuit. Specifically, they happen when a $Z$ error occurs on a $Z$ ancilla qubit due to its CNOT gate in timestep 2. Through the CNOT interactions in timesteps 3 and 4, the $Z$ error is propagated to two data qubits, $d_1$ and $d_4$. When qubit $x_1$ interacts with qubit $d_1$ in timestep 4, the $Z$ error has already propagated to qubit $d_1$, so it is detected in round $i$. However, by the time the $Z$ error propagates to qubit $d_4$ (timestep 4), qubit $x_2$ has already interacted with qubit $d_4$ for round $i$, so it does not detect the error. Qubit $x_2$ is only able to detect the $Z$ error one round later in round $i+1$.

\begin{figure}
    \centering
    \includegraphics[width=0.6\columnwidth]{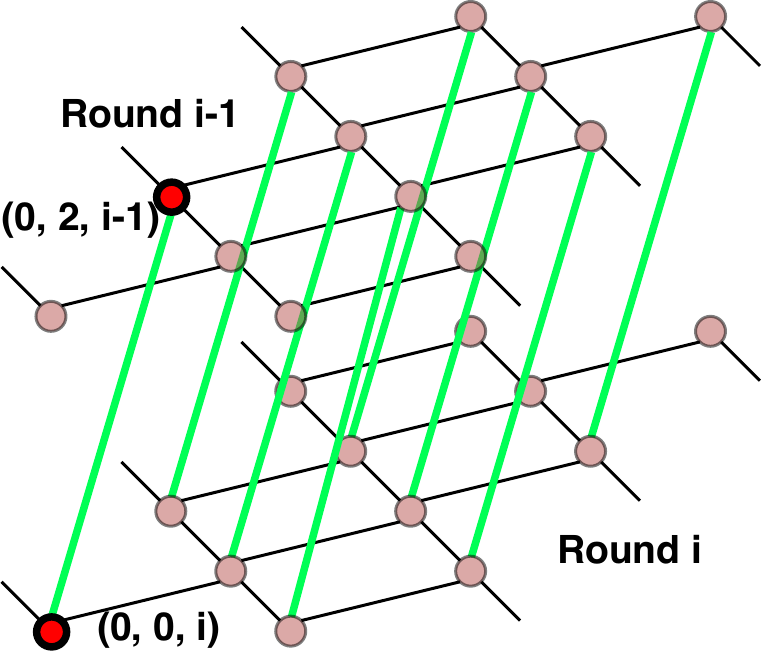}
    \caption{Full coverage for hook spacetime-like errors in the predecoding subgraph mapped to a single pipeline stage, \textbf{H}.}
    \label{fig:hook-coverage}
\end{figure}

For Pinball to achieve full coverage of hook errors, we require predecoding primitives which consider pairs of syndromes across consecutive rounds which are separated vertically by \emph{two} rows of data qubits (Fig. \ref{fig:hook-coverage}). Then, each of these predecoding primitives must correct along the chain of two data qubits between them. It does not matter whether the pair of qubits $d_1$ and $d_4$ or qubits $d_2$ and $d_3$ are corrected, since these corrections differ by a stabilizer.

\subsection{Comparison with Clique} \label{cmp_clique}
Pinball's coverage substantially improves compared to the current state-of-the-art cryogenic predecoder, Clique \cite{ravi2023better}. As indicated in Tab. \ref{tab:coverage-quality}, Clique includes robust handling of space-like errors, limited handling of time-like errors, and no logic for handling spacetime-like errors. As such, Clique fails to account for many types of error propagation in the syndrome measurement circuit. Conversely, through detailed circuit-level analysis, we have derived the predecoding primitives necessary to maintain strong coverage of all three error classes. As we show in Sec. \ref{sec:evaluation}, these design improvements enable Pinball to significantly outperform Clique in realistic noise settings.

\begin{table}
\caption{Predecoding primitive checks: Pinball vs. Clique.}
\centering
\begin{tabular}{|c|c|c|}
\hline
\textbf{Predecoding Primitive Check} & \textbf{Clique}     & \textbf{Pinball} \\ \hline
Bulk Space-like Errors            &  \greentext{Good}               & \greentext{Good}             \\ \hline
Edge Space-like Errors            & \yellowtext{Okay} & \greentext{Good}             \\ \hline
Time-like Errors                  & \orangetext{Poor}          & \greentext{Good}             \\ \hline
Spacetime-like Errors             & \redtext{Absent}        & \greentext{Good}             \\ \hline
\end{tabular}
\label{tab:coverage-quality}
\end{table}

\subsection{Resolving Predecoding Primitive Conflicts} \label{sec:predecoding-conflicts}
\begin{figure}
    \centering
    \includegraphics[width=0.9\columnwidth]{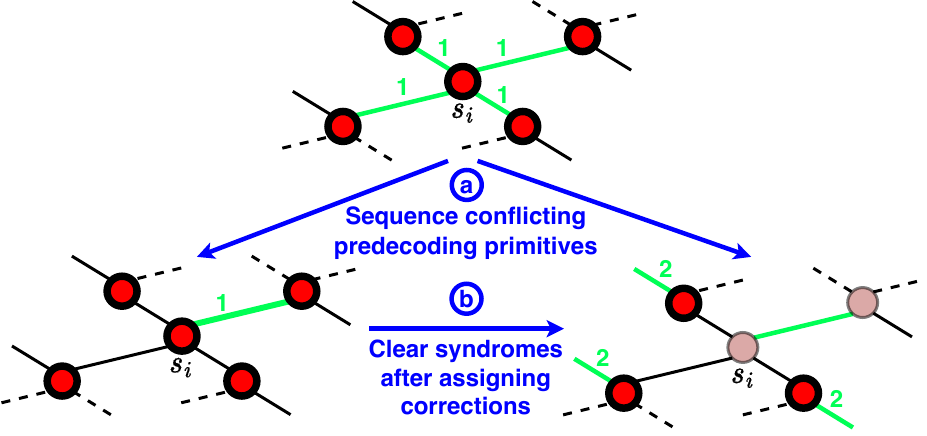}
    \caption{By sequencing conflicting predecoding primitives and clearing active syndromes (highlighted) after assigning corrections, race conditions are resolved.}
    \label{fig:conflicting-primitive}
\end{figure}

Since every syndrome in the predecoding subgraph has degree greater than one, an active syndrome, $s_i$, could have multiple neighboring syndromes which are also active. 
Consider the top scenario in Fig \ref{fig:conflicting-primitive}: as we have described it thus far, Pinball would naively assign corrections to all four edges between $s_i$ and its active neighbors. 
However, if these four errors had actually occurred, $s_i$'s error parity would have been even, so it never would have been active - thus, the correction is wrong. This contradiction is due to a race condition between the four predecoding primitives which share $s_i$. If these primitives are allowed to execute in parallel, they quadruple-count $s_i$ by mapping it to four predecoding subgraph edges.

To avoid this, two requirements must be met. First, predecoding primitives sharing a syndrome between them must be executed sequentially, one after the other; we refer to such predecoding primitives as \textit{conflicting}. This resolves the race condition by enforcing an ordering between them. Second, to avoid re-counting the same syndrome for multiple errors, each predecoding primitive must clear its pair of active syndromes after mapping them to their corresponding predecoding subgraph edge. These two steps are shown at the bottom of Fig \ref{fig:conflicting-primitive}. Clearing syndromes after assigning corrections also enables very lightweight complex detection logic: if any active syndromes remain after executing all predecoding primitives, the error pattern must be complex, and the \textit{original, unmodified syndromes} are propagated to the RT decoder to generate corrections.

\section{Hardware Design} \label{sec:pinball-impl}
Having defined Pinball's algorithm, we now translate it into a lightweight, low-power hardware design. We first discuss the hardware for the predecoding primitive and then expand to a full design covering the entire predecoding subgraph.

\subsection{Predecoding Primitive Logic} \label{sec:predecoding-primitive-logic}
\begin{figure*}
    \centering
    \includegraphics[width=\textwidth]{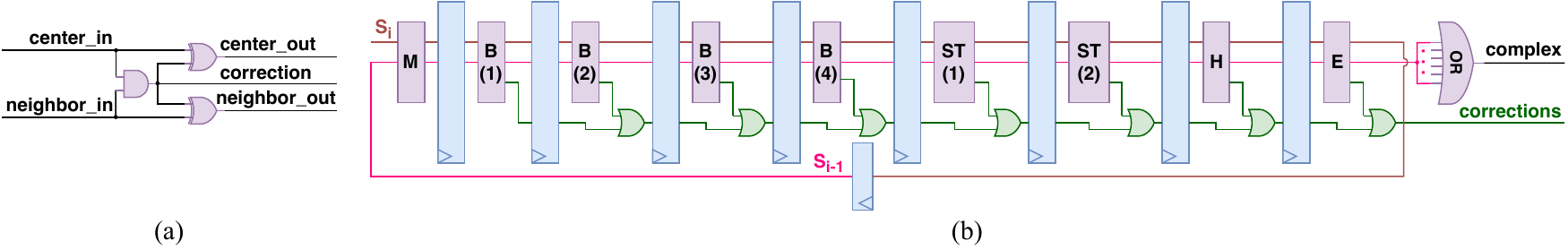}
    \caption{The hardware design for Pinball. (a) The two-level combinational logic implementing a predecoding primitive. (b) Pinball is organized into a pipeline, where each pipeline stage processes a disjoint group of predecoding primitives.}
    \label{fig:pinball-hw}
\end{figure*}

A predecoding primitive is responsible for checking a pair of syndromes in the predecoding subgraph, and if they are both active, assigning a correction to the data qubit(s) corresponding to the edge between them. To avoid double-counting active syndromes, it must also clear its syndromes after assigning a correction. The hardware implementing this logic is organized as a two-level combinational circuit as shown in Fig. \ref{fig:pinball-hw}a.

For each pair of syndromes, we call one the ``center" and the other the ``neighbor". The first-level AND gate indicates if both syndromes are active, and if so, a (set of) correction bit(s) are sent to the location(s) in a $d \times d$ buffer corresponding to the data qubits that should be corrected. In the second level, the AND gate's output is fed into two XOR gates to conditionally clear the pair of active syndromes if a correction is assigned.

\subsection{Pipeline Design} \label{sec:pipeline-design}
To avoid race conditions in assigning corrections, conflicting predecoder primitives must be executed sequentially. To orchestrate this in hardware, we organize Pinball into a sequence of \underline{nine pipeline stages} (Fig. \ref{fig:pinball-hw}b) where each stage executes a disjoint set of conflict-free primitives in parallel. We now discuss how we derive the nine disjoint sets in each stage.

Beginning with space-like errors, we observe that each syndrome within the bulk of the decoding graph has four neighbors. We refer to the corresponding edges as \textit{bulk space-like errors}. Per bulk syndrome, none of the four predecoding primitives covering its edges can be executed in parallel, and so four separate pipeline stages, \textbf{B(1)}-\textbf{B(4)}, are needed.

Space-like errors at the edges of the lattice activate only a single edge syndrome, so we use predecoding primitives with these syndromes as center and \textit{artificial} neighbors. Artificial syndromes are always set to active, allowing Pinball to assign corrections conditioned only on the values of the isolated edge syndromes. These edge space-like errors are handled in a separate stage, \textbf{E}, for reasons described in Sec. \ref{sec:pipeline-benefits}.

In Fig. \ref{fig:time-like-coverage}, it is clear that, for time-like errors, none of the predecoding primitives conflict. Thus, they can all be executed in parallel within a single pipeline stage which we call \textbf{M}.

Finally, recall that spacetime-like errors are organized into two distinct sets: one for errors on a single data qubit and the other, hook errors, for errors on two qubits. From Figs. \ref{fig:spacetime-like-coverage} and \ref{fig:hook-coverage}, most syndromes participate in decoding primitives within both sets, so we handle each set in separate stages.

For single-qubit spacetime-like errors, most syndromes participate in two predecoding primitives, one which leans to the bottom-left from round $i-1$ to round $i$ (solid lines in Fig. \ref{fig:spacetime-like-coverage}) and the other which leans to the bottom-right (dashed lines in Fig. \ref{fig:spacetime-like-coverage}). We separate these groups into two pipeline stages called \textbf{ST(1)} and \textbf{ST(2)}. Within the set of hook spacetime-like errors, no predecoding primitives share syndromes, so they can all be executed in parallel in a single stage which we call \textbf{H}.

In the final pipeline stage, if any syndromes in $S_{i-1}$ (and $S_i$ in the last round of the $d$-round block) remain active, Pinball was clearly unable to account for all errors, and second-level decoding is required. An OR-reduction tree detects this case, and its \texttt{complex} output triggers offloading of the unmodified $d$ syndrome rounds to RT. Importantly, our selection of disjoint groups allows Pinball to achieve \textit{constant predecoding latency per syndrome round independent of code distance.}

\subsection{Pipelining Benefits} \label{sec:pipeline-benefits}
The use of pipelining in Pinball has three primary benefits compared to prior cryogenic predecoding work. 

\circled{1} First, it allows for the decoding problem to be progressively simplified \textit{in situ} by clearing syndromes in each stage as Pinball generates new corrections. Consequently, full coverage of the decoding graph can be achieved with minimal decoding hardware. Each edge is covered only once and in a single stage throughout the pipeline, so our design features no redundant hardware. This lies in contrast to Clique \cite{ravi2023better} which redundantly covers each space-like edge twice in order to provide a full set of corrections in one pass. This increased efficiency reduces the additional hardware overhead incurred by Pinball to cover the new spacetime-like edges.

\circled{2} Second, pipelining provides a natural enforcing of order between decoding primitive checks, allowing us to prioritize certain checks over others by placing them earlier in the pipeline. We describe two heuristics for ordering checks. First, decoding primitive groups covering the most common errors should be placed earlier in the pipeline to minimize the likelihood that their associated syndromes are paired with other active syndromes as part of different checks for less common errors. Second, the decoding primitives for edge space-like errors should be placed at the end of the pipeline, since they clear only one syndrome at a time. By doing so, we account for more syndromes through fewer potential errors which has a higher likelihood of being correct.

To properly order checks, we simulated $10^5$ shots of $d$-round syndrome blocks to obtain a frequency distribution of different error classes in the \texttt{SI1000} noise model (see Sec. \ref{sec:methodology}). In Tab. \ref{tab:error-dist}, time-like and space-like errors dominate, suggesting their checks should be prioritized. Empirically, we observed best performance when checking time-like errors first. 

\begin{table}
\caption{Distribution of length-1 errors for $d=11$ and $p=10^{-3}$ in the \texttt{SI1000} noise model.}
\centering
\begin{tabular}{|c|c|}
\hline
        \textbf{Error Type}       & \textbf{Frequency (\%)} \\ \hline
Time-like Errors                    & 48.02              \\ \hline
Space-like Errors                     & 41.65              \\ \hline
Spacetime-like Errors (single qubit) & 7.69              \\ \hline
Hook Errors                          & 2.63             \\ \hline
\end{tabular}
\label{tab:error-dist}
\end{table}

\circled{3} Finally, since there are no control dependencies between Pinball's pipeline stages, it is a fully modular design providing complete freedom to reorder/re-prioritize certain decoding primitive checks. This maybe be useful if different error sources start to dominate over others.

\subsection{Pinball Failure Modes} \label{sec:pinball-failure}
\begin{figure}
    \centering
    \includegraphics[width=\columnwidth]{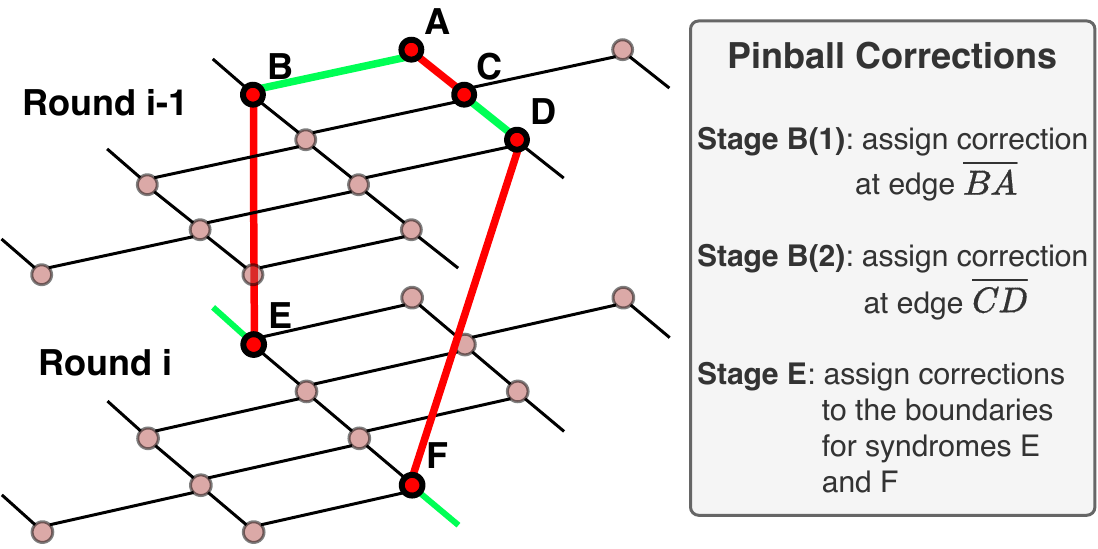}
    \caption{A set of errors which Pinball would both fail to decode correctly and flag as complex. Red (green) edges correspond to errors (corrections).}
    \label{fig:pinball_failure}
    \vspace{-0.25cm}
\end{figure}

While Pinball's heuristics yield high predecoding performance (Sec. \ref{sec:evaluation}), there are inevitably some scenarios in which Pinball both fails to provide an accurate set of corrections \textit{and} fails to detect that the decoding is complex, leading to a logical error. Fig. \ref{fig:pinball_failure} shows an illustrative example. First, suppose the ordering of bulk space-like checks happens to check edges $\overline{BA}$ and $\overline{CD}$ in stages B(1) and B(2), respectively, before edge $\overline{CA}$ is checked in stage B(4). Consequently, Pinball assigns corrections to the wrong space-like edges, and syndrome pairs $BE$ and $DF$ are broken up, isolating syndromes $E$ and $F$.

Normally, isolated syndromes raise Pinball's complex flag, but critically, since they both reside at the edges of the decoding graph, Pinball can clear both in stage E. More generally, Pinball's local heuristics are vulnerable in scenarios where it breaks up a pair of active syndromes, and the remaining isolated syndrome lies at the edge of the decoding graph.

\subsection{Predecoding Imbalance: Energy Saving Opportunity}
\label{sec:low-power}

From Sec. \ref{sec:decoding-algorithms}, recall that the decoding rate must be at least as high as the rate of syndrome generation. This sets a most stringent per-round decoding latency budget of 1\,$\mu$s for superconducting qubit systems. Since the predecoder introduces additional end-to-end latency to the decoding hierarchy, this must be accounted for when determining the operating frequency of the hardware. We can allocate a conservative 10\% of the per-round decoding latency budget (100 ns) for predecoding, ensuring that enough time is always reserved for syndrome transmission to RT and second-level decoding. At high code distances, this is much less than 1\% of the $d$ $\mu s$ required to generate syndrome data, so we can expect Pinball's latency overhead to be negligible.

\begin{figure}[t]
    \centering
    \includegraphics[width=\columnwidth]{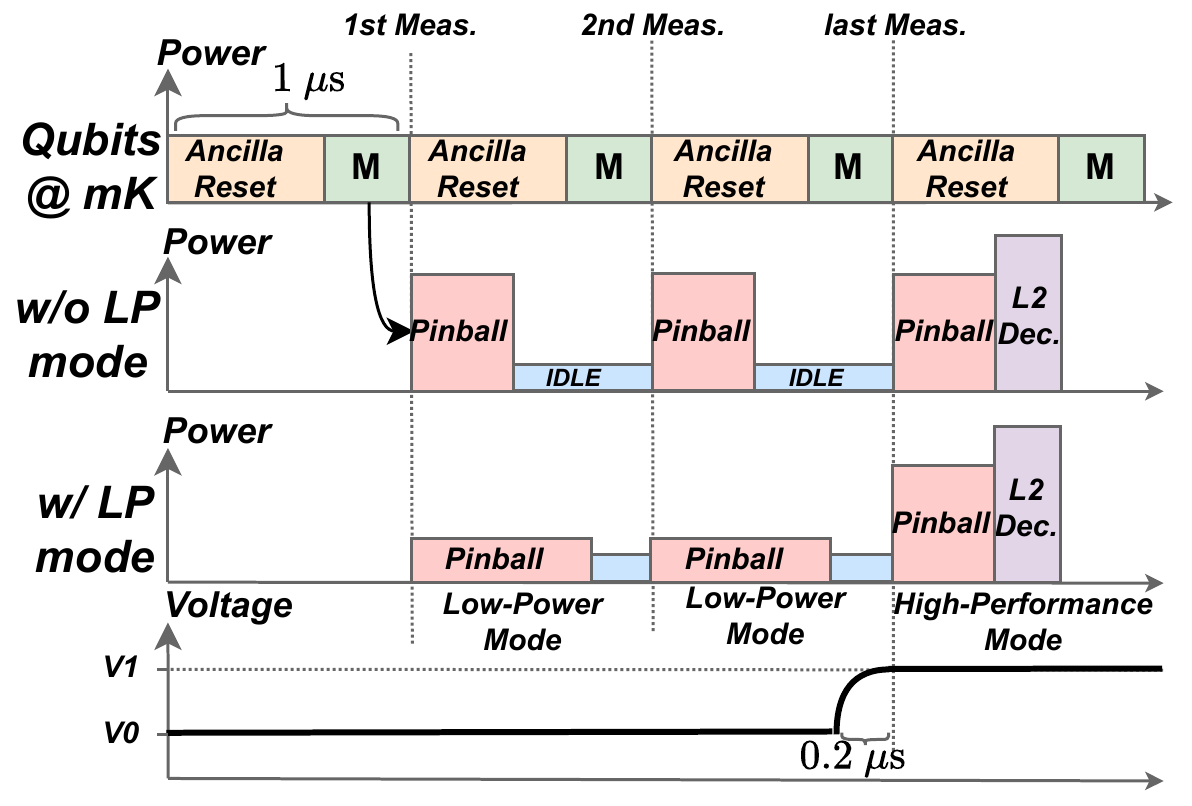}
    \caption{Workload-aware performance scaling by selecting from different supplies and body biases. Voltage switching occurs 200\,ns prior to the end of the penultimate round.}
    \label{fig:dvfs}
\end{figure}

Naively, we would use this 100 ns budget across all $d$ rounds (middle row in Fig. \ref{fig:dvfs}) since the same predecoding is performed in each round. However, notice that \emph{the predecoding latency impacts the critical path only in the last ($d$-th) round when data may be transmitted to the second-level decoder}. Thus, we identify a significant per-round latency constraint imbalance in the predecoding process: in the first $d-1$ rounds, predecoding can utilize the full 1\,$\mu$s budget, whereas it must operate within the 100\,ns budget in the last ($d$-th) round.

We exploit this imbalance to lower power and energy consumption through `workload'-aware performance scaling. For the first $d-1$ rounds, we use a ``low-power (LP) mode", extending processing time while operating at a lower voltage and frequency, whereas in the last round, we use a ``high-performance (HP) mode", ramping up to higher voltage and frequency to decrease processing time. Fig. \ref{fig:dvfs} details the proposed workload-aware performance scaling scheme.

We employ a power multiplexer to switch between supply voltages \cite{fan2024enhancing} and voltage multiplexers to switch between corresponding body biases. On-chip overhead is limited to only a simple state machine to control the transitions, a small LUT to store voltage-frequency pairs, and a small bank of multiplexing transistors.  Transistors can be moderately sized to minimize both power and area overhead, and voltage-frequency pairs can be calibrated during post-silicon testing to compensate for process variation and minimize energy. Additionally, power multiplexing supports fast switching times, enabling the voltage to settle within 200\,ns \cite{fan2024enhancing} after which the clock frequency can be elevated. As shown in Fig. \ref{fig:dvfs}, with the workload-aware performance scaling scheme enabled, Pinball occupies at most 800\,ns of the 1\,$\mu$s window, leaving 200\,ns for the power multiplexers to settle before the $d$-th round.

CMOS threshold voltages increase significantly at cryogenic temperatures, limiting the operable supply voltage headroom; however, forward body biasing can effectively reduce threshold voltages~\cite{overwater2023cryogenic}. The 22\,nm FDSOI technology supports a wide body biasing range, offering designers fine-grained control of threshold voltages and performance~\cite{flatresse2013ultra}. To characterize the impact of supply voltage on the maximum operating frequency at 4\,K in the 22\,nm FDSOI technology, we constructed a ring oscillator consisting of 301 inverter gates and simulated it under various supply voltages and body biases (VBN for NMOS and VBP for PMOS). The resulting frequency and power data are shown in Tab. \ref{tab:ro_exp}. At supply voltages above 0.54\,V, the ring oscillator oscillates successfully without any body bias adjustment. At 0.5\,V, forward body biasing becomes necessary to lower the threshold voltages and maintain oscillation. As the supply voltage decreases further, both the operating frequency and power consumption drop significantly. Below 0.47\,V, the ring oscillator fails to oscillate regardless of body biasing. We refer to Tab.~\ref{tab:ro_exp} in Sec. \ref{sec:evaluation} to estimate the energy savings achieved by our workload-aware performance scaling scheme.

\begin{table}
\caption{Frequency and power of a ring oscillator at 4\,K across varying supply and body bias voltages.}
\centering
\small
\begin{tabular}{|c|c|c|c|c|c|c|}
\hline
\textbf{Supply (V)} & \textbf{0.8} & \textbf{0.6} & \textbf{0.54} & \textbf{0.5} & \textbf{0.48} & \textbf{$<$0.47} \\
\hline
VBN (V) & 0 & 0 & 0 & 0.2 & 0.3 &  \\
\cline{1-6}
VBP (V) & 0 & 0 & 0 & -0.3 & -0.3 & Fails to \\
\cline{1-6}
Freq. (MHz) & 355.9 & 158.5 & 93 & 72.1 & 53.2 & Oscillate \\
\cline{1-6}
Power ($\mu$W) & 184.5 & 16.1 & 2.6 & 1.6 & 0.98 & \\
\hline
\end{tabular}
\label{tab:ro_exp}
\end{table}
\section{Methodology} \label{sec:methodology}

\subsection{Noise Model} \label{sec:noise-model}
In our evaluations, we use a circuit-level noise model which captures measurement errors, single-qubit and two-qubit gate errors, reset errors, and qubit idling errors. Within circuit-level noise, the error rates of each error source can be scaled by a base physical error rate, $p$. We simulate $p$ in the range $10^{-4}-10^{-2}$ to capture performance between higher-end error rates for present-day hardware ($10^{-2}$) and realistic, future hardware error rates ($10^{-4}$). Tab. \ref{tab:noise-model} summarizes the \texttt{SI1000} noise model \cite{gidney2021fault} we use, a realistic noise model inspired by superconducting hardware.

\begin{table}
    \caption{Superconducting hardware noise model \cite{gidney2021fault}.}
    \centering
    \begin{tabular}{|c|c|c|c|c|c|c|}
        \hline
        \textbf{Error Source} & \textbf{Meas.} & \textbf{Reset} & \textbf{1Q}   & \textbf{2Q} & \textbf{Idle} & \textbf{Resonator Idle} \\ \hline
        \textbf{Error Rate}   & $5p$                 & $2p$           & $\frac{p}{10}$ & $p$          & $\frac{p}{10}$ & $2p$ \\ \hline
    \end{tabular}
    \label{tab:noise-model}
\end{table}

\subsection{Decoding Simulation Framework} \label{sec:decoding-sim-frame}
\label{sec:method-sw}
We run Monte Carlo simulations of decoding on a single logical qubit, varying code distance with odd values between $d=3-21$. We use Stim \cite{gidney2021stim} to generate $d+1$ rounds of random error patterns per simulation shot. Corresponding syndromes are sent to the decoder. Since $X$ and $Z$ errors are handled identically, we only simulate $Z$ errors, as is typical in prior QEC work \cite{das2022afs, ravi2023better, liyanage2023scalable, wu2023fusion, vittal2023astrea, holmes2020nisq+, alavisamani2024promatch}. 

Per round, we first supply syndromes to Pinball. If it cannot find a full set of corrections, the round is marked complex. If any round with a $d$-round block is marked complex, the block is passed to Pymatching \cite{higgott2025sparse}, a Python minimum-weight-perfect matching (MWPM) second-level decoder. In either case, we check if the final corrections cause a logical error. In parallel to Pinball, we also simulate Clique \cite{ravi2023better} and Promatch \cite{alavisamani2024promatch} to compare with state-of-the-art cryogenic and RT predecoders. For all experiments using a predecoder, we use Pymatching as the L2 decoder. Finally, since \cite{alavisamani2024promatch} found that running Promatch in parallel with Astrea-G \cite{vittal2023astrea}, a greedy L2 decoder, improved LER, we also compare Pinball to this ensemble configuration labeled Promatch$||$Astrea-G. Both the predecoder/decoder hierarchies as well as Astrea-G are given a total of $d$ $\mu$s to complete decoding.

We evaluate Pinball's \textit{L1 coverage}, the percentage of $d$-round blocks where none of the rounds were marked complex, and its \textit{L1 accuracy}, the percentage of non-complex blocks decoded correctly, and compare them with Clique's over $10^7$ $d$-round blocks of syndromes. The former is a proxy for syndrome bandwidth reduction, and the latter is a proxy for logical error rate. For all predecoders, we simulate up to $10^9$ syndrome blocks to determine logical error rate.

\subsection{Hardware Characterization}
\label{sec:method-hw}
We implement Pinball in 22\,nm FDSOI technology to measure its power, performance, and area. Standard cells were re-characterized with a commercial 4\,K device model to accurately capture cryogenic behavior, after which synthesis, placement, and routing were performed. The resulting netlist was simulated with delay back-annotation, and its activity vectors enable precise per-error-rate 4\,K power estimates.

In HP mode, Pinball operates at 0.8\,V and 100\,MHz, while in LP mode, it operates at 0.48\,V and 12.5\,MHz, meeting the 800 ns latency over nine clock cycles. Using Tab. \ref{tab:ro_exp}, we scale power and energy according to $P \propto CV^2f$ and $E \propto CV^2$, respectively, where $C$ is the effective switching capacitance, $V$ the supply voltage, and $f$ the operating frequency.

We compare Pinball's energy consumption with both SFQ and 22 nm FDSOI implementations of Clique. Since the on-chip cost for power multiplexing and body biasing is very low, we omit power multiplexing and body biasing when estimating Pinball's energy consumption. As a baseline, we also consider a system without a cryogenic predecoder where syndrome data is always sent to RT. Two factors contribute to Pinball's energy reduction. First, through its L1 coverage, it reduces the transmission energy used to transmit syndromes to RT. Second, it reduces energy at 4\,K by operating in LP mode for $d-1$ rounds, only switching to HP mode in the last round. In our assessment of these impacts, we use the 2.46\,pJ/bit energy efficiency of a recently-proposed cryogenic transmitter \cite{fakkel2023cryo} to estimate transmission energy.
\begin{figure}
    \centering
    \includegraphics[width=\columnwidth]{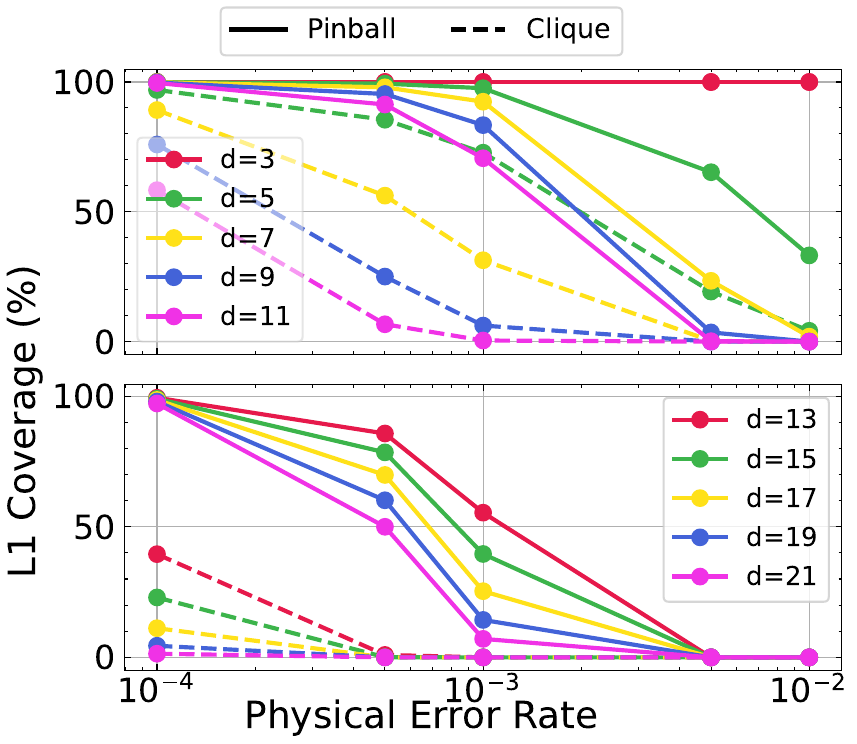}
    \caption{L1 coverage comparison between Pinball (solid) and Clique (dashed) showing that Pinball decodes a higher percentage of syndromes within the 4 K stage of the dilution refrigerator.}
    \label{fig:coverage}
\end{figure}

\section{Evaluation} \label{sec:evaluation}

\subsection{System-Level Cryogenic Predecoding Benefits} \label{sec:cryo-perf}
Cryogenic predecoding aims to maximize decoding within the cryogenic domain (L1 coverage) without loss of accuracy (L1 accuracy), thereby reducing syndrome bandwidth while maintaining low LER. We evaluate how effectively Pinball achieves this compared to the state-of-the-art Clique \cite{ravi2023better}.

\noindent
\textbf{L1 Coverage and Bandwidth Reduction}: As shown in Fig. \ref{fig:coverage}, Pinball consistently achieves higher coverage than Clique, and the gap increases with code distance. For near-term error rates ($p=10^{-3}$), Pinball increases coverage up to 77.22\% at $d=9$, reducing syndrome transmission to RT by 5.65$\times$ relative to Clique. For long-term error rates ($p=10^{-4}$), Pinball increases coverage up to 95.75\% at $d=21$, giving an additional 34.72$\times$ transmission reduction. Note that marginal gains in coverage still give striking system-level performance benefits: both Pinball and Clique have high coverage ($> 89\%$) at $p=10^{-4}$ and $d=7$, but Pinball's 10.72\% increase gives 124.33$\times$ higher transmission reduction.

Pinball's coverage improvements are due to three factors. First, and most importantly, Clique fails to consider syndrome pairs across space \textit{and} time, so it cannot correct the significant fraction (Tab. \ref{tab:error-dist}) of spacetime-like errors. Second, Clique handles edge space-like errors before measurement errors, ruling out matching pairs of active syndromes too quickly. Third, for time-like errors, while Pinball clears the syndromes at both endpoints, Clique only suppresses the one in the oldest round. Due to the latter two factors, Clique breaks up more syndrome pairs within the same and across different rounds, leaving more isolated syndromes to be marked complex.

\begin{figure}
    \centering
    \includegraphics[width=\columnwidth,trim={0.2cm 0.3cm 0.3cm 0.1cm},clip]{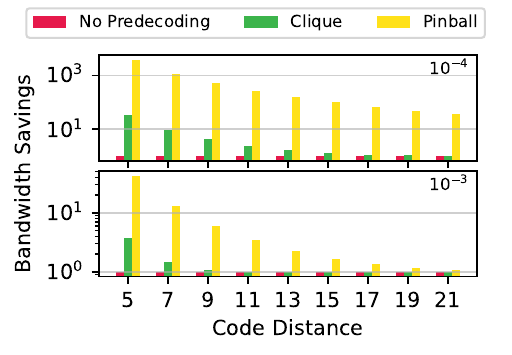}
    \caption{Bandwidth savings comparisons between Pinball and Clique, normalized to a system which does not use predecoding. Pinball achieves higher bandwidth savings than Clique due to increased L1 coverage.}
    \label{fig:bw-reduction}
\end{figure}

Based on these coverage improvements, we show the increase in syndrome bandwidth savings achieved by Pinball in Fig. \ref{fig:bw-reduction}. Clearly, independent of the underlying physical error rate, $p$, Pinball achieves exceptional bandwidth savings; at $d=3$, Pinball achieves full coverage and at $d=5$ $p=10^{-4}$, reduces total bandwidth by a maximum $3780.72\times$. In the worst case ($d=21$ and $p=10^{-3}$), $1.08\times$ reduction is still achieved.

\begin{figure}
    \centering
    \includegraphics[width=\columnwidth]{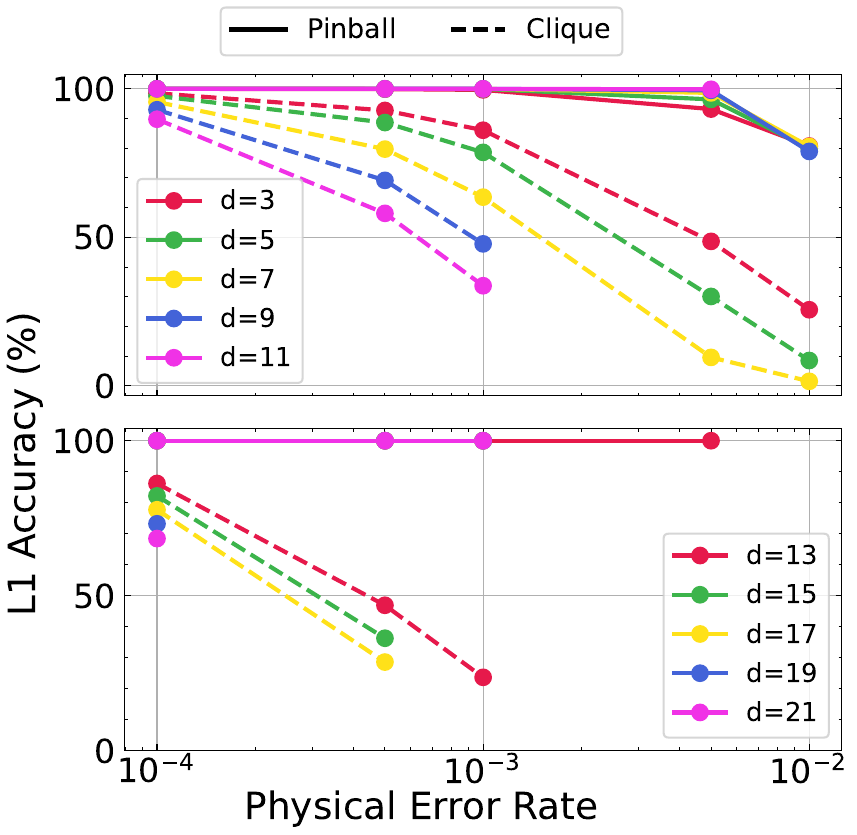}
    \caption{L1 accuracy comparison between Pinball (solid) and Clique (dashed) showing that Pinball decodes syndromes with higher accuracy than Clique. Note: some data points omitted in due to insufficient L1 coverage.}
    \label{fig:accuracy}
\end{figure}

\noindent
\textbf{L1 Accuracy and Logical Error Rate:} In Fig. \ref{fig:accuracy}, we compare Pinball's L1 accuracy to Clique. Again, Pinball consistently achieves higher accuracy, and the gap increases with code distance and physical error rate. At $p=10^{-3}$, Pinball increases accuracy up to 4.23$\times$ at $d=13$. For $d> 13$, Pinball maintains 100\% while Clique's drops to 0\%. At $p=10^{-4}$, Pinball still gives up to 1.46$\times$ accuracy increase at $d=21$.

\begin{figure}
    \centering
    \includegraphics[width=\columnwidth]{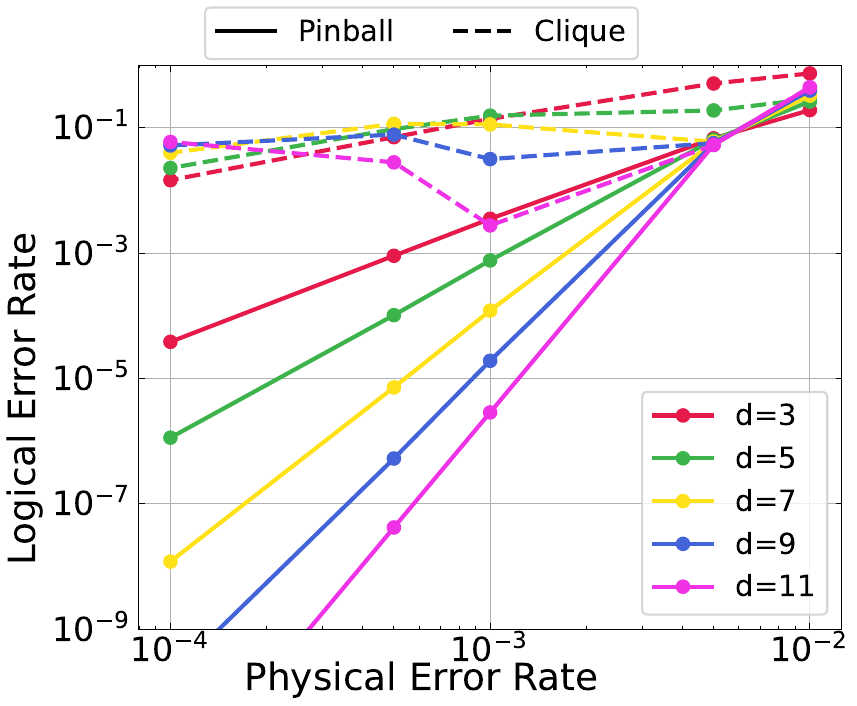}
    \vspace{-0.5cm}
    \caption{Logical error rate comparison between Pinball and Clique.}
    \label{fig:pinball_vs_clique_ler}
\end{figure}

Pinball's improved L1 accuracy translates into radically different LER performance in Fig. \ref{fig:pinball_vs_clique_ler}. As opposed to Clique, using Pinball as the L1 cryogenic predecoder exponentially reduces LER as physical error rate lowers and code distance increases. At $p=5\cdot 10^{-4}$ and $d=11$, \textit{Pinball improves LER by nearly six orders of magnitude compared to Clique}. Clique only achieves comparable LER at high code distances and physical error rates, since in this regime, it offloads most decoding to Pymatching at L2. Again, seemingly marginal gains yield considerably different system-level outcomes. At $p=10^{-4}$ and $d=7$, Pinball's ``small" L1 accuracy increase of 4.42\% reduces LER by nearly seven orders of magnitude.

These results are especially noteworthy when considered together with coverage: \textit{not only does Pinball handle more decoding scenarios than Clique, it handles this greater number with higher accuracy.} On the surface, this result seems counter-intuitive. As Clique's coverage decreases, it is becoming more selective with which syndromes it chooses to handle itself. However, despite being more selective, its accuracy continues to decrease. This is not the case for Pinball; even as its coverage decreases, it consistently maintains high accuracy, suggesting it is more robust against false positives (complex syndromes which it misclassifies as simple) than Clique.

In summary, while Clique was instrumental in laying foundations for cryogenic predecoding, our results indicate it is less likely to provide direct benefits in real-world quantum computing systems. Pinball is the first cryogenic predecoder providing system-level benefits under realistic circuit noise.

\subsection{Comparison with RT Predecoding} \label{sec:rt-predecoding-comp}
\begin{figure}
    \centering
    \includegraphics[width=\columnwidth]{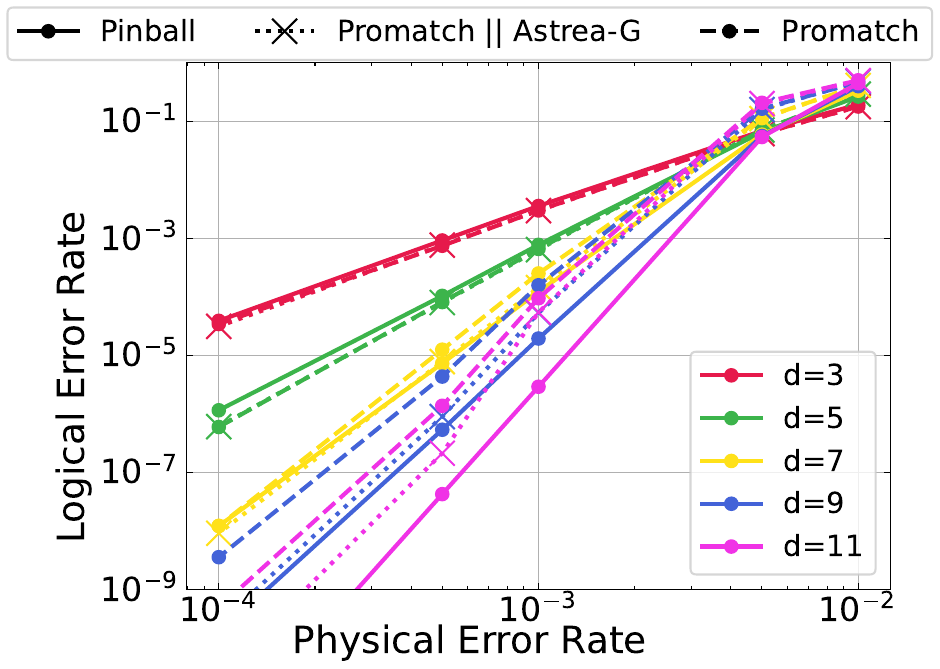}
    \vspace{-0.5cm}
    \caption{Logical error rate comparison between Pinball, Promatch, and the ensemble Promatch$||$Astrea-G \cite{alavisamani2024promatch}.}
    \label{fig:pinball_vs_promatch_ler}
\end{figure}

Next, we contrast Pinball to a recent state-of-the-art RT predecoder, Promatch \cite{alavisamani2024promatch}. As mentioned in Sec. \ref{sec:predecoding}, an ideal decoding hierarchy could feature both cryogenic and RT predecoders. To study their respective predecoding capabilities, we compare the LER of Pinball to Promatch in Fig. \ref{fig:pinball_vs_promatch_ler} when both are independently integrated with Pymatching.

Remarkably, we find that, despite stricter power and area constraints limiting Pinball's complexity, it still achieves a lower LER than both Promatch and Promatch$||$Astrea-G for $d \geq 7$, and the performance gap widens as code distance increases; at $p=5 \cdot 10^{-4}$ and $d=11$, we observe a 32.38$\times$ (5$\times$) reduction in LER compared to Promatch (Promatch$||$Astrea-G). We attribute the improvement over Promatch to the different prioritization schemes used in each design: Promatch prioritizes corrections which avoid creating isolated syndromes, whereas Pinball also accounts for the frequencies of different error classes in the ordering of its pipeline stages. Further, Astrea-G often struggles to generate matchings within its allocated latency budget due to denser syndromes under SI1000 noise, limiting the ensemble configuration's benefits. Its results may be improved by increasing its latency budget/operating frequency, but this is orthogonal to our work. Importantly, this result shows that, in a decoding hierarchy with  Pinball and Promatch, the system's LER would be limited by Promatch's performance, not Pinball's. For Pinball's full benefits, it could be used with higher-accuracy RT predecoders in the future.

\subsection{Cryogenic Predecoding Impacts on Logical Error Rate} \label{sec:predecoding-impacts}
So far, we have compared systems with different predecoders. While Pinball improves LER over these, it must also match the performance of a system without predecoding to ensure its bandwidth gains do not compromise fidelity.

\begin{figure}
    \centering
    \includegraphics[width=\columnwidth]{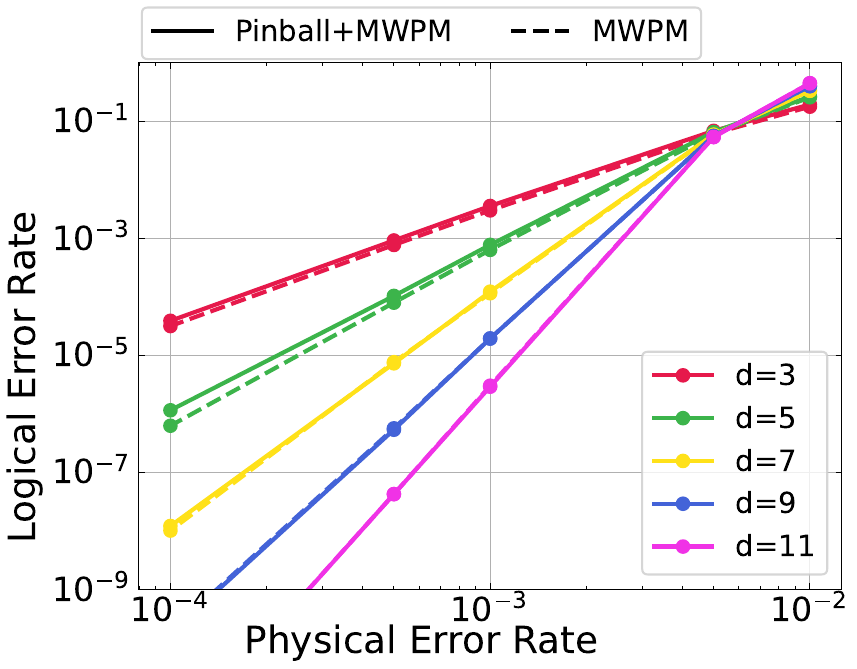}
    \vspace{-0.5cm}
    \caption{Logical error rate comparison between a system using Pinball as an L1 predecoder and a system with no predecoder.}
    \label{fig:pinball+mwpm_vs_mwpm}
\end{figure}

To confirm this, in Fig. \ref{fig:pinball+mwpm_vs_mwpm}, we evaluate LER for two configurations: one using Pinball at L1 and Pymatching at L2 and one using only Pymatching at L2. Since Pymatching provides state-of-the-art decoding accuracy, we see a slight increase in LER when using Pinball for low $d$ and physical error rates ($d \leq7, p <10^{-3}$). However, beyond $d=7$, the system using Pinball closes the gap to achieve near parity in LER, demonstrating that using Pinball maintains the required QEC performance.

\subsection{Hardware Overheads} \label{sec:hardware-overheads}
\begin{figure}
    \centering
    \includegraphics[width=\columnwidth]{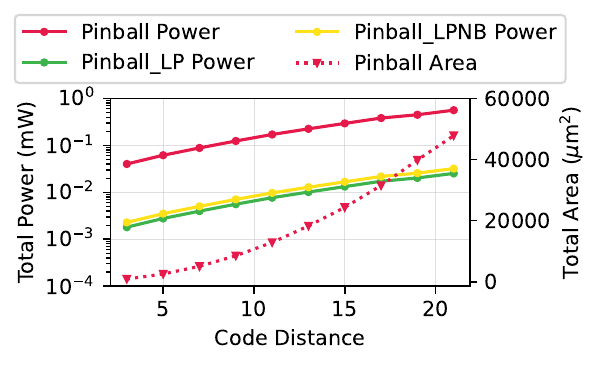}
    \vspace{-0.7cm}
    \caption{Area and power analysis of Pinball's HP and LP mode with (Pinball\_LP) and without (Pinball\_LPNB) body biasing.}
    \label{fig:pinball-area-power}
\end{figure}

Fig. \ref{fig:pinball-area-power} presents the power and area consumption of Pinball (HP mode), Pinball\_LPNB (LP mode, no body biasing) and Pinball\_LP (LP mode, body biasing), evaluated for $p=10^{-3}$. In area, Pinball only consumes $< 0.05\,mm^2$, even at the highest $d=21$. Pinball consumes between 0.04\,mW and 0.56\,mW of power in HP mode. Based on Sec. \ref{sec:method-hw}, Pinball\_LPNB can operate at 0.54\,V supply voltage, while Pinball\_LP can operate even lower at 0.48\,V. Both variants can operate at a 12.5\,MHz frequency which satisfies the latency constraint of 800\,ns (Sec. \ref{sec:low-power}). Thus, Pinball\_LPNB and Pinball\_LP achieve $17.5\times$ and $22.2\times$ power reductions, respectively, compared to HP mode.

Assuming a 1.5\,W cooling budget at 4\,K~\cite{krinner2019engineering}, and conservatively using worst-case power (i.e, HP mode during the $d$-th syndrome round), at $p=10^{-3}$, we support enough Pinball instances within a single dilution refrigerator to simultaneously decode 37,313 logical qubits at $d=3$ and 2,668 logical qubits at $d=21$, showing our design reaches targets well within the early fault-tolerant quantum computing (eFTQC) regime \cite{katabarwa2024early}. 

\subsection{Energy Savings}
\begin{figure}
    \centering
    \includegraphics[width=\columnwidth,trim={0.25cm 0.2cm 0.2cm 0.2cm},clip]{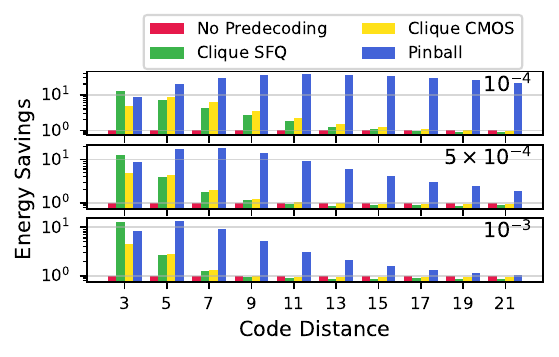}
    \caption{Total energy savings combines the benefits of predecoding and LP mode operation in the first $d-1$ rounds. Results are shown for physical error rates between $10^{-4}$ (top) and $10^{-3}$ (bottom).}
    \label{fig:energy}
\end{figure}

We evaluate the energy of Pinball, CMOS/SFQ Clique, and the baseline system, including predecoder and 4\,K-to-RT transmission costs. As shown in Fig.~\ref{fig:energy}, the predecoder's hardware overhead grows with code distance and error rate, reducing its benefit. Still, Pinball’s higher coverage allows it to consistently outperform CMOS and SFQ Clique for $d > 3$.

For $p = 10^{-3}$, corresponding to the eFTQC regime, energy savings range from $1.05\times$-$13.35\times$. eFTQC devices containing $\sim$$10^4$ physical qubits will likely use code distances below 11 to support 50-100 logical qubits. At these lower code distances, Pinball provides $3.15\times$-$13.35\times$ energy savings. 

As error rates decrease, higher L1 coverage improves energy savings. Pinball's savings range from $1.92$-$18.57\times$ for $p = 5 \cdot 10^{-4}$ and from $8.61$-$37.05\times$ for $p=10^{-4}$. Thus, as hardware improves, Pinball remains scalable and effective for the higher code distances needed in long-term FTQC. These estimates assume only syndrome payloads are transmitted, underestimating actual transmission costs. In practice, protocols like Riverlane’s QEC interface (QECi) \cite{riverlane2024qeci} use 64-bit packets with 32-bit headers, doubling transmission cost. Accounting for this boosts Pinball’s savings to $1.06$–$22.77\times$ at $p = 10^{-3}$, and up to $67.4\times$ at $p = 10^{-4}$.

\section{Discussion \& Related Work} \label{sec:discussion}
\textbf{QEC Codes:} Although we focus on the surface code, many works have explored the general family of quantum low-density parity check codes \cite{delfosse2022toward, gu2023efficient, stein2025hetec, vittal2024flag, sahay2022decoder, gidney2023new}. We anticipate predecoding to be equally valuable for them.

\textbf{Second-Level Decoders:} While we analyzed Pinball with an MWPM-based second-level decoder, many other decoders exist \cite{delfosse2021almost, liyanage2023scalable, barber2025real, wu2023fusion, vittal2023astrea, das2022afs, das2022lilliput}. Our design can be interfaced with any of them. Some work has investigated full-fledged cryogenic decoders \cite{holmes2020nisq+, ueno2022qulatis, ueno2021qecool}, but strict 4\,K constraints preclude scalability without sacrificing accuracy.

\textbf{Predecoders:} Prior works have used neural networks \cite{chamberland2023techniques} and greedy heuristics \cite{delfosse2020hierarchical, alavisamani2024promatch, ravi2023better, smith2023local} to simplify decoding. Only Clique \cite{ravi2023better} considers cryogenic operation. Relative to state-of-the-art RT and cryogenic predecoders, Promatch \cite{alavisamani2024promatch} and Clique \cite{ravi2023better}, Pinball achieves much lower LER. 

\textbf{Cryo-CMOS and SFQ Logic:} Advances have been made in both cryo-CMOS \cite{bardin201929, chakraborty2022cryo, underwood2024using, van2020scalable, park2021fully, kossel202440, hinderling2024flip, bersano2023quantum, frank2023low, bardin2019design, zou2021frequency} and SFQ \cite{kim2024fault, choi2024supercore, holmes2020nisq+,jokar2022digiq,ishida2020supernpu,kashima202164} technologies as well as modeling tools \cite{byun2022xqsim, min2022cryowire, min2023qisim}. As they mature, they may permit deeper decoding hierarchies at intermediate temperature stages (e.g., 77\,K) inside the dilution refrigerator.

\textbf{Qubit Control and Readout Power:} 
Current cryo-CMOS control and readout electronics~\cite{bardin201929, chakraborty2022cryo, yoo2023isscc, kang202334, guo202429, huang2025cryogenic} draw too much power, requiring higher efficiency to stay within the 4\,K budget. Once trimmed, syndrome transmission power will remain prominent, motivating bandwidth reduction techniques like predecoding. Equally important, the predecoding hardware  must add only minimally to the power overhead. These constraints drive the optimizations presented.
\section{Conclusion} \label{sec:conclusion}
In this paper, we presented Pinball, a novel cryogenic predecoder tailored to surface code QEC decoding under circuit-level noise. We analyzed error generation and propagation at the circuit-level to increase predecoder performance, and using an architecture co-optimized with cryo-CMOS technology, effectively managed increased predecoding complexity. 
Pinball is the first scalable cryogenic predecoder capable of operating effectively under noise present in real quantum systems.

\section*{Acknowledgements}
This material is based upon work supported by the U.S. Department of Energy, Office of Science, Office of Advanced Scientific Computing Research, Accelerated Research in Quantum Computing under Award Number DE-SC0025633. This research used resources of the National Energy Research Scientific Computing Center, a DOE Office of Science User Facility supported by the Office of Science of the U.S. Department of Energy under Contract No. DE-AC02-05CH11231 using NERSC award ASCR-ERCAP0033197. The authors also thank Semiwise Ltd., UK for access to the cryo-CMOS PDK used for hardware evaluation in this work.


\balance
\bibliographystyle{IEEEtranS}
\bibliography{refs}

\end{document}